%% file: paper.tex
\documentclass[11pt]{article}

\usepackage{graphicx}
\usepackage{url}
\usepackage{amssymb}
\usepackage{amsfonts}
\usepackage{amsmath}

\usepackage{algorithm}
\usepackage{algpseudocode} 
\usepackage{multicol}
\usepackage{multirow}
\usepackage{diagbox}
\usepackage{booktabs}
\usepackage{threeparttable}
\usepackage{tablefootnote}
\usepackage{bm}
\usepackage{svg}
\usepackage{dirtytalk}

\usepackage{listings}
\usepackage{xcolor}
\usepackage{tcolorbox}
\usepackage{stfloats}

\definecolor{Maroon}{RGB}{128, 0, 0} 

\newcommand{\e}[1]{e_{#1}}

\lstdefinestyle{sqlstyle}{
    language=SQL,
    basicstyle=\ttfamily\footnotesize,
    keywords={SELECT, FROM, WHERE, GROUP, BY, AVG, as},
    keywordstyle=\color{blue},
    commentstyle=\color{darkgray},
    stringstyle=\color{red},
    breaklines=true,
    morekeywords={},
}

\newtcolorbox{mytakeaway}{
    colback=gray!10,
    colframe=white,
    coltitle=white,
    fonttitle=\bfseries,
    left=1mm,
    right=1mm,
    top=1mm,
    bottom=1mm
}

\newtcolorbox{mytakeaway_main}{
    colback=red!10,
    colframe=white,
    coltitle=white,
    fonttitle=\bfseries,
    left=1mm,
    right=1mm,
    top=1mm,
    bottom=1mm
}

\newcommand{\centeredItalic}[1]
{
    \begin{center}
        \textit{#1}
    \end{center}
}

\def\QED{$\square$}

\newtheorem{example}{Example}
\newtheorem{theorem}{Theorem}
\newtheorem{definition}{Definition}

\title{The Selection Problem in Multi-Query Optimization:\\ a Comprehensive Survey}

\author{Sergey~Zinchenko and Denis~Ponomaryov\\~\\
Novosibirsk State University\\  Institute of Informatics Systems, Novosibirsk, Russia\\
s.zinchenko@alumni.nsu.ru, ponom@iis.nsk.su}
\date{}

\begin{document}

\maketitle

\begin{abstract}
View materialization, index selection, and plan caching are well-known techniques for optimization of query processing in database systems. The essence of these tasks is to select and save a subset of the most useful candidates (views/indexes/plans) for reuse within given space/ time budget constraints. In this paper, we propose a unified view on these selection problems. We make a detailed analysis of the root causes of their complexity and summarize techniques to address them. Our survey provides a modern classification of selection algorithms known in the literature, including the latest ones based on Machine Learning. We provide a ground for reuse of the selection techniques between different optimization scenarios and highlight challenges and promising directions in the field. Based on our analysis we derive a method to exponentially accelerate some of the state-of-the-art selection algorithms.
\end{abstract}

\input{chapters_introduction}

\input{chapters_related_work}

\input{chapters_preliminaries}

\input{chapters_preparation}

\input{chapters_algorithms}

\input{chapters_applying}

\input{chapters_challenges}

\input{chapters_conclusion}

\bibliographystyle{ACM-Reference-Format}
\bibliography{paper}

\end{document}

%% file: chapters_introduction.tex
\section{Introduction}
\label{chapter:introduction}

With the growing data storage and analysis demands, \textit{Data Warehouses (DWH)} became increasingly widespread providing an unified access to data from a large number of heterogeneous sources. To mitigate the costs of configuring, maintaining, and scaling DB systems, platforms based on \textit{Database-as-a-Service} (\textit{DBaaS}) are now being widely implemented. Due to a typically large number of similar requests to service based DB systems \textit{Multi-Query Optimization (MQO)} proves useful as a technique that aims at finding and reusing common computations for a more efficient workload execution. Savings achieved by re-use are typically called \textit{benefit} in the literature. In general, the task is to find candidate computations for re-use which provide the highest benefit, while respecting the constraints on the available resources (e.g., disk space or computing time). This task can be subdivided into three \textit{orthogonal} problems: 1) discovering common computations between queries, 2) \textit{selecting} the most useful ones, and 3) making an optimal plan for their re-use. In this paper, we focus on the second problem, i.e., the problem of selection.

One obtains different instances of this problem depending on the type of common candidate computations considered and the range of possible actions over the selected candidates. For example, in DWH scenarios, free disk space can be used to save (\textit{materialize}) common data (\textit{views}) \cite{widom1995research}. The pre-computed data can then be read from the disk instead of computing it from scratch, which can speed up query execution by several orders of magnitude. The selection problem in this scenario is typically called \textit{View Selection Problem (VSP)}, which is to identify a set of views that gives the highest benefit for a workload and fits the storage and maintenance budgets. Execution of a workload can also be accelerated by creating indexes that make access to data faster. In general, the \textit{Index Selection Problem (ISP)} is similar to VSP because an index can be considered as a special case of a single-table, projection-only materialized view \cite{agrawal2000automated}. MQO is also employed in the context of stored procedures and analytical reports, because due to the relatively small data and high processing time for reports, the latency of report generation can be reduced by storing candidate computations for re-use in the memory. In the literature, the selection problem for this scenario is referred to as \textit{Query (Result) Caching} and it is very similar to VSP. Another way to speed up queries is to reduce planning time. This can be achieved by caching good plans and by reusing them for similar queries. The problem of selecting the most useful plans in such scenarios is known as \textit{Plan Caching}.  

These instances of the selection problem have certain specifics in different scenarios. In this paper we note however that in fact \textbf{selection algorithms are agnostic to the nature of candidates they manipulate with\footnote{only the way objects are represented and benefits are computed is important}, and thus, they can be reused between different optimization scenarios under similar constraints}. Motivated by this observation, we formulate a generalized Candidate Selection Problem that abstracts away from the nature of candidates: they can be views, indexes, cached data, or even plans. 

\textbf{Our contribution} can be summarized as follows:

\begin{enumerate}    
\item we introduce a modern classification of selection algorithms, including the recent ones using Machine Learning;
    
\item we propose a general framework of Candidate Selection which allows for reusing ideas and techniques between different instances of selection problems including View/Index Selection and Query/Plan Caching;

\item we make a detailed analysis of the root causes of the complexity of selection problems and summarize techniques to address them;

\item based on our analysis we derive a method to exponentially accelerate some of the state-of-the-art selection algorithms;
    
\item we highlight challenges, open questions, and promising directions in the development of selection algorithms for Multi-Query Optimization.
\end{enumerate}

The paper is organized as follows. In Section \ref{chapter:related work}, we review related literature including surveys on similar topics and highlight the main differences with our work. In Section \ref{chapter:preliminaries}, we begin our exposition with examples of the main issues related to selection procedures and formulate the general Candidate Selection Problem. Then in Section \ref{chapter:preparation} we introduce techniques to resolve these issues. We continue our study in Section \ref{chapter:algorithms} with a classification of selection algorithms, emphasizing the main techniques that can be reused to solve the Candidate Selection Problem (and thus, instances thereof). In Section \ref{chapter:applying}, on the basis of our analysis we derive a technique to exponentially speed up two SotA View Selection algorithms. Finally, in Section \ref{chapter:challenges} we describe open questions, challenges, and promising directions for future research and in Section \ref{chapter:conclusion} we conclude.

%% file: chapters_related_work.tex
\section{Related Work}
\label{chapter:related work}

Most related to our work is the survey paper \cite{mami2012survey} from 2012 which provides a classification of algorithms for View Selection. Recently some novel algorithms have been proposed, not covered in \cite{mami2012survey}, which present a line of research related to Machine Learning. We review these algorithms in detail in our paper. Also, unlike \cite{mami2012survey}, we make a detailed analysis of issues behind the proposed algorithms and explain the reasons for the design decisions in these algorithms. We believe that this study could facilitate the development of better solutions. We also propose a general framework within which one can use the previously proposed selection algorithms interchangeably in different selection scenarios. 

We now list the topics that are intentionally not covered in this paper. One of the very first surveys on View Selection was focused on the problem of choosing a view definition language and self-maintaining a set of views \cite{gupta1995maintenance}. This topic, as well as the problem of finding an optimal way to use materialized views (\textit{Query Answering Using Views} \cite{halevy2001answering}), is relevant for building efficient solutions, but it is not focused on the selection itself and thus, not addressed in our survey. Our exposition also does not cover techniques from Data Mining {\cite{sohrabi2016materialized}, Constraint Programming {\cite{mami2011modeling}, and Game Theory {\cite{azgomi2018game}, which did not attract much interest in the context of selection problems. Also, we do not consider techniques for updating selected candidates, as well as the relationship of the selection problem to the topics such as stream data processing, approximate query processing, etc. We recommend \cite{chirkova2012materialized} as a starting point on these topics. For an introduction to the Index Selection Problem, we recommend 
 \cite{chaudhuri2004index}. We also do not touch questions of coupling computation caching with other techniques such as query execution scheduling and pipelining, which are studied in \cite{diwan2006scheduling}. We also mention that details on the problem of Plan Caching for a template of parameterized queries can be found in \cite{stoyanovich2008reoptsmart,hulgeri2002parametric,hulgeri2003anipqo,ghosh2002plan}. In the literature, these topics are referred to as \textit{Parametric Query Optimization}.

%% file: chapters_preliminaries.tex
\section{Preliminaries}
\label{chapter:preliminaries}

We begin our exposition with an analysis of the principal issues related to selection problems in Multi-Query Optmization. We introduce the required terminology, then we formulate the Candidate Selection Problem as a unified view on selection tasks. Then we summarize results related to the computational complexity of this problem.

\subsection{Multi-Query Optimization} 

\centeredItalic{What are the goals and main problems of Multi-Query Optimization?}

\begin{figure}
    \centering
    \includegraphics[width=0.8\linewidth,keepaspectratio]{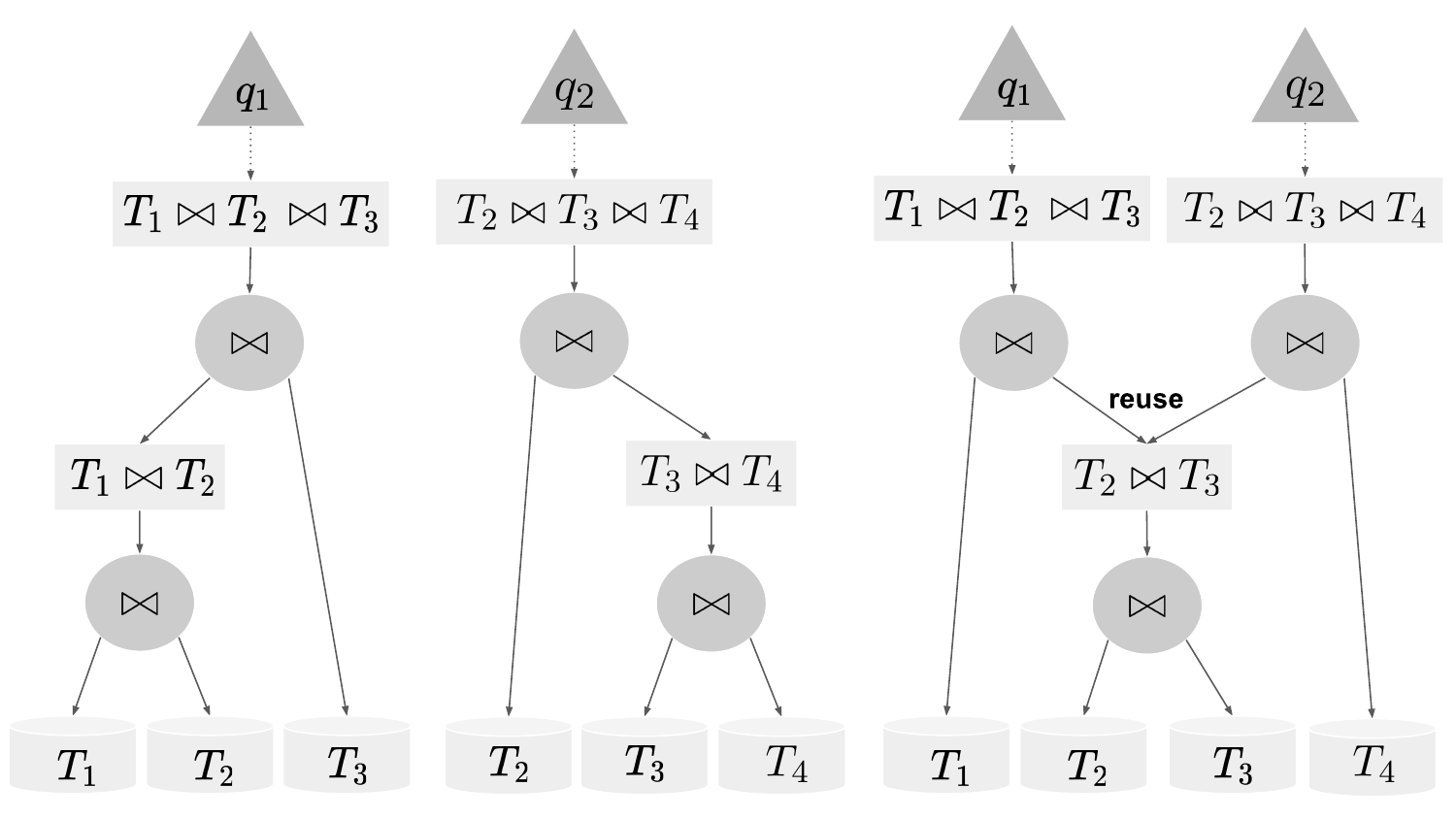}
    \caption{To speed up query execution it may be useful to \textit{find} and \textit{reuse} common computations. Although the computation of $T_2 \bowtie  T_3$ is not a part of an optimal plan of either query, it may be optimal for executing both queries together.    On the left: optimal plan for $q_1$. In the middle: optimal plan for $q_2$. On the right: optimal plan for the whole workload.}
  \label{fig:reuse}
\end{figure}

\noindent Consider a workload $Q$ and assume that besides the base relations in a database we have some precomputed information (e.g., views, indexes, or cached query plans) denoted as $C$. We can use them for processing queries from $Q$, and thus, it holds that the execution plan for an individual query $q \in Q$ generally depends on $C$, so we denote it as $\mathcal{P}_C(\{q\})$. The union of plans for all $q \in Q$ is called \textit{execution plan for workload} $Q$ and we denote it as $\mathcal{P}_C(Q)$. 

The Multi-Query Optimization Problem (MQO) is to compute a set $C$ such that $\mathcal{P}_C(Q)$ is optimal for executing $Q$. This problem is \textit{fundamentally more complex than individual query optimization}, because it introduces the option of reusing items from $C$ between queries, as shown below in an example of VSP:

\begin{example}
    \label{ex:reuse}(Ex. 1.1 from \cite{roy2000efficient})
    Let workload $Q$ consist of queries $q_1 = T_1 \bowtie T_2 \bowtie T_3$, $q_2 = T_2 \bowtie T_3 \bowtie T_4$ and let $\mathcal{P}(\{q_1\}) = (T_1 \bowtie T_2) \bowtie T_3$ and $\mathcal{P}(\{q_2\}) = T_2 \bowtie (T_3 \bowtie T_4)$ be their \textit{individually} optimal plans, respectively. Although these plans are optimal for each query separately, it can be the case that by \textit{reusing} $C = \{T_2 \bowtie T_3\}$ one obtains an optimal plan for the entire workload $Q$. That is, using the join sequences $T_1 \bowtie (T_2 \bowtie T_3)$ and $(T_2 \bowtie T_3) \bowtie T_4$ gives a total latency for both $q_1$ and $q_2$ lower than the sum of the latencies obtained by using $\mathcal{P}(\{q_1\})$ and $\mathcal{P}(\{q_2\})$. An illustration is given in  Figure \ref{fig:reuse}.
\end{example}

    \noindent The challenge in MQO is to find such reuse cases efficiently. Moreover, since the resources to store and maintain such computations are limited, we arrive at the task of selecting the most useful ones.

\begin{mytakeaway}
    The goal of MQO is to jointly optimize a series of queries by (a) \textit{identifying} computations that can be efficiently reused and (b) \textit{saving} the most useful ones within a storage/maintenance budget.
\end{mytakeaway}

\subsection{Representation of Candidates} 

\begin{figure}
  \centering
  \includegraphics[width=0.8\linewidth,keepaspectratio]{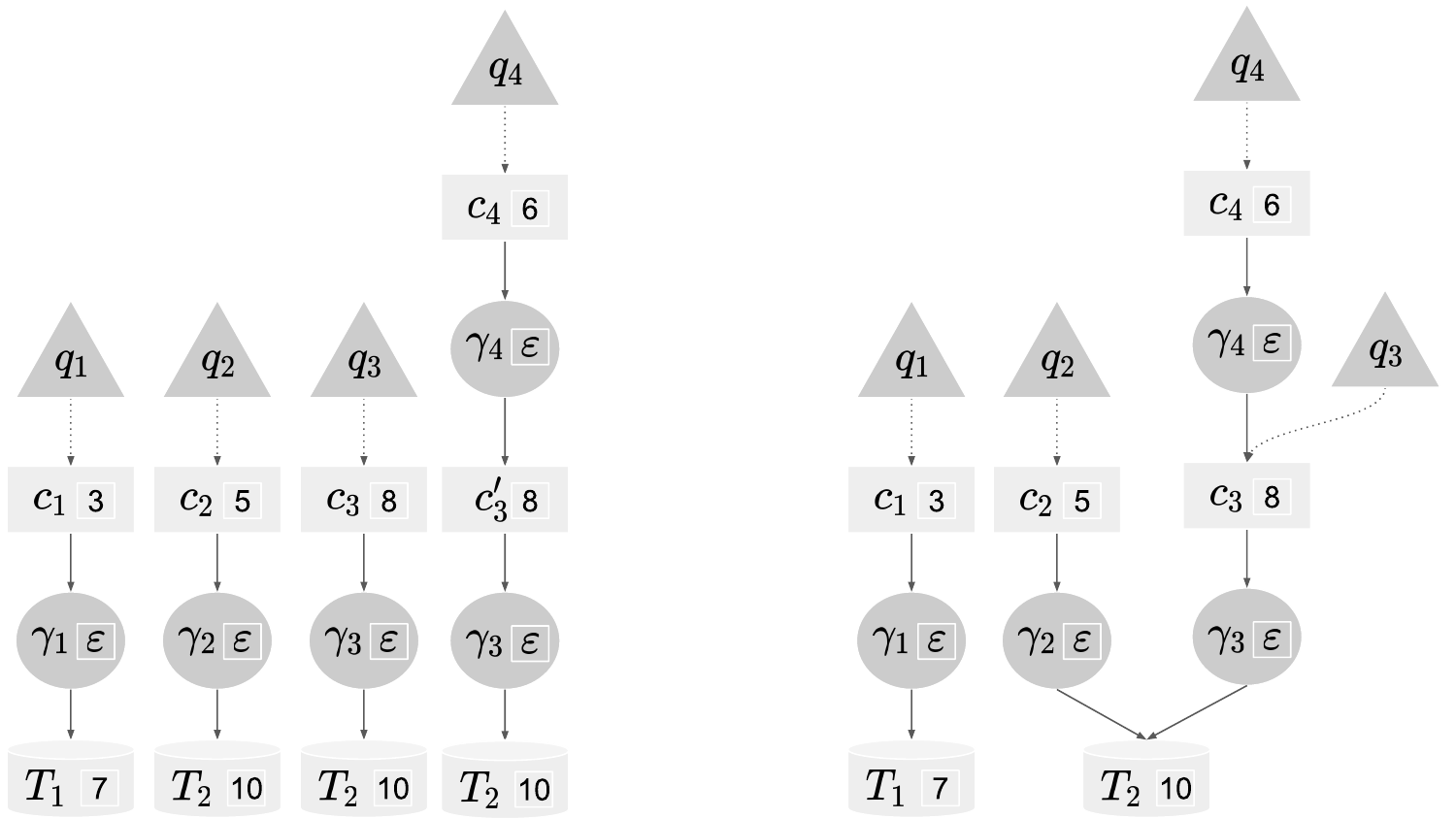}
  \caption{To discover options for computation reuse, a procedure of merging several expression trees into a expression forest can be applied.  On the left: representation of the workload in the form of expression trees. On the right: the result of merging them into a expression forest $\mathcal{F}$. Eq-nodes ($c_i$) are shown in rectangles, op-nodes ($\gamma_i$) are shown in circles. Data sizes and execution times for operations are shown in small boxes inside these figures\protect\footnotemark. Triangle-shaped nodes are used to depict which data every query $q_i$ needs.} 
  \label{fig:expression_tree}
\end{figure}

\footnotetext{we assume that the read time for data is the same as its size, and the data is permanently saved to disk between executions of operations}

\label{section:preliminaries_representation} 

\centeredItalic{Why representation of candidates matters?}

Typically each query is  represented in the form of an \textit{expression tree} which is built from execution plan. An expression tree is a directed bipartite acyclic graph, which represents required data, operations over it, and the result. A more formal definition is provided below.

\begin{definition}
    \label{def:expression_tree}
    An expression tree is a pair $\langle V_{op} \sqcup V_{eq}, E \rangle$, where $V_{op}$ is a set of operation nodes (op-nodes, for short), $V_{eq}$ is a set of data nodes (known in the literature as equivalence nodes or eq-nodes, for short\footnote{due to the fact that for each data node there can be several computation paths, which give \textit{equivalent} results}), and $E$ is a set of directed edges, which can be of two types. The first edge type ($v_{eq} \rightarrow v_{op}$) indicates that operation $v_{op}$  is applied to data $v_{eq}$. The second type ($v_{op} \rightarrow v_{eq}$) indicates that $v_{op}$ must be executed to retrieve data $v_{eq}$. 
\end{definition}

Then expression trees obtained for individual queries are joined into an \textit{expression forest} by merging common nodes which reflects the possibility of reusing computations. An example of a merging procedure is shown below.

\begin{example}
    \label{ex:merge}
    Let workload $Q$ consist of the following four queries
    \begin{lstlisting}[style=sqlstyle]
q1: SELECT week, AVG(price) as avg_price 
    FROM T1 GROUP BY week;
q2: SELECT week, AVG(price) as avg_price 
    FROM T2 GROUP BY week;
q3: SELECT week, day, AVG(price) as avg_price 
    FROM T2 GROUP BY week, day;
q4: SELECT week, AVG(avg_price) as macro_avg_price 
    FROM (
        SELECT week, day, AVG(price) as avg_price
        FROM T2 GROUP BY week, day
    ) as weekly_price 
    GROUP BY week;
    \end{lstlisting}  
    \noindent the optimal plans $\mathcal{P}(\{q_1\}), \ldots , \mathcal{P}(\{q_4\})$ for which are shown on the left in Fig. \ref{fig:expression_tree}. Then plans $\mathcal{P}(\{q_2\}), \mathcal{P}(\{q_3\})$ and $\mathcal{P}(\{q_4\})$ are merged due to the common read of table $T_2$. Also, since the aggregation by $\langle week, day \rangle$ in queries $q_3$ and $q_4$ is the same, the data of $c_3$ and $c_3'$ is the same and these nodes are merged too. 
\end{example}

\noindent After that candidates for reuse are searched among eq-nodes of the resulting expression forest. As each eq-node corresponds to the result of the execution of some subexpression, the search space in this approach is also called \textit{subexpression space}.

We note that this approach has a significant \textit{drawback}. Since an expression forest is built over  optimal plans for individual queries, some candidates for building a plan optimal for the entire workload can be \textit{missed} (as shown in Example \ref{ex:reuse}). In order to obtain an optimal solution, one needs to represent a query as an expression tree in a way that \textit{takes into account alternative computation paths}. To achieve this, one can use graphs of a special type, which we consider further in Section \ref{subchapter:query_representation}. However, having all possible expression trees available may not suffice in general. As shown by Example 1 in \cite{chirkova2002formal}, the most optimal solution may contain a candidate which \textit{is not a subexpression} of any of the queries in a workload even if we consider \textit{all alternatives}. 

\begin{mytakeaway}
    Choosing an appropriate \textbf{representation} is an essential step in solving the selection problem as it defines the search space for candidates, and as a consequence, the quality of the best solution in it. 
    %\textbf{the optimality of selection}. 
\end{mytakeaway}

\subsection{Benefit of Candidates} 

\centeredItalic{What is a benefit and why is it hard to compute?}

\noindent As we see, the first step is to define a search space for candidates, from which the most beneficial candidates are selected. The overall benefit, denoted as $\mathfrak{B}(C)$, for a subset $C$ of candidates relative to a workload $Q$ is typically defined as savings in execution time, i.e., the difference between the execution time of $Q$ without the selected candidates, $T_{\varnothing}(Q)$, and the execution time of $Q$ with the selected candidates, $T_C(Q)$:
\[
\mathfrak{B}(C) = T_{\varnothing}(Q) - T_C(Q),
\]
where $T_S(Q)$ denotes the execution time of $Q$ when using plan $\mathcal{P}_S(Q)$. The \textit{benefit} $\mathcal{B}_c$ of a candidate $c$ is defined as the execution time saving obtained by reusing $c$ in the optimal plan $\mathcal{P}_C(Q)$. This allows us, in the case of expression trees without alternative execution paths (i.e., AND-DAGs, see definition in Chapter \ref{subchapter:query_representation}), to perform the following decomposition of the overall benefit:
\begin{equation} \label{eq:benefit_decomposition}
\mathfrak{B}(C) = \sum_{c \in C} \mathcal{B}_{c}(\textcolor{Maroon}{\textbf{C}}) = \sum_{c \in C} \underline{\mathcal{B}_{c}'} \cdot n\_reuses_c(\textcolor{Maroon}{\textbf{C}}),
\end{equation}
where $\mathcal{B}_{c}'$ is the speedup obtained by reusing candidate $c$ once, i.e., the difference between its execution cost and its reuse cost (e.g., reread), and $n\_reuses_c$ is the number of times candidate $c$ is reused in the optimal plan $\mathcal{P}_C(Q)$. This decomposition allows us to isolate the part $\mathcal{B}_{c}'$, which is \underline{independent} of the selected set of candidates. It is important to note that this benefit $\mathcal{B}_c$ \textcolor{Maroon}{still depends} on other elements in $C$, because an optimal plan depends on the whole set of candidates available for reuse:

\begin{example}
    \label{ex:benefit}
    Consider the workload from Example \ref{ex:merge} depicted in Figure \ref{fig:expression_tree}. Suppose that the data retrieval process consists of a) reading all operands relevant to the corresponding operation and b) executing it. Then in plan $\mathcal{P}_{\{c_3\}}(Q)$ we use $c_3$ to answer \textit{both} queries $q_3$ and $q_4$, so the benefit $\mathcal{B}_{c_3}(\{c_3\})$ equals 
    $$
    \mathfrak{B}_{c_3}' \cdot n\_reuses_{c_3}(\{c_3\}) = [ex\_cost_{c_3} - reuse\_cost_{c_3}] \cdot 2 = [(10 + \varepsilon) - 8] \cdot 2 = [2 + \varepsilon] \cdot 2.
    $$ 
    Instead of reading table $T_2$ and executing $\gamma_3$ in $10 + \varepsilon$ time, we just read $c_3$ \textit{twice}, each time spending $8$ units of time. However, in plan $\mathcal{P}_{\{c_3, c_4\}}(Q)$ the benefit $\mathcal{B}_{c_3}(\{c_3, c_4\})$ equals $\boldsymbol{1} \cdot [2 + \varepsilon]$ ($< \boldsymbol{2} \cdot [2 + \varepsilon]$), as we use $c_3$ only to execute $q_3$. This demonstrates that the benefit of $c_3$ \textit{is decreased} when $c_4$ is selected. At the same time, the benefit of candidate $c_4$ equals
    $$
    \mathcal{B}_{c_4}(\{c_3, c_4\}) = [ex\_cost_{c_4} - reuse\_cost_{c_4}] \cdot 1 = [(10 + \varepsilon + \varepsilon) - 6] \cdot 1 = [4 + 2\varepsilon] \cdot 1,
    $$
    so the total benefit \( \mathfrak{B}_{\{c_3, c_4\}} \) is increased and satisfies Equation~\ref{eq:benefit_decomposition}:
    \[
    \begin{aligned}
    \mathfrak{B}(\{c_3, c_4\}) = & \ T_{\emptyset}(Q) - T_{\{c_3, c_4\}}(Q) \\
    = & \ [T_{\emptyset}(\{q_1\}) - T_{\{c_3, c_4\}}(\{q_1\})] + [T_{\emptyset}(\{q_2\}) - T_{\{c_3, c_4\}}(\{q_2\})] + [T_{\emptyset}(\{q_3\}) - T_{\{c_3, c_4\}}(\{q_3\})] \\
    = & \ [(10 + \varepsilon) - (10 + \varepsilon)] + [(10 + \varepsilon) - 8] + [(10 + 2\varepsilon) - 6] \\
    = & \ \mathcal{B}_{c_1}' \cdot 0 + \mathcal{B}_{c_2}' \cdot 1 + \mathcal{B}_{c_3}' \cdot 1 \\
    = & \ \sum_{c \in \{c_3, c_4\}}\mathcal{B}_{c}' \cdot n\_reuses_c(\{c_3, c_4\}).
    \end{aligned}
    \]

\end{example}

\noindent Executing a workload workload with all possible sets of selected candidates is infeasible. Therefore, in practice, benefits are calculated by using some estimates which leads us to the next problem: estimates for operation latency and data size for nodes in an expression forest may be highly \textit{inaccurate} (they are typically obtained from a database optimizer). Moreover the estimation errors are multiplied if the same operation is used on several computation paths simultaneously. Since it is problematic to obtain an optimal solution with wrong estimates, the problem of accurate benefit estimation becomes \textit{crucial}. We consider different ways to approach this problem in Section \ref{subchapter:benefit_estimation}. 

\begin{mytakeaway}
    The typical understanding of candidate's benefit is how much it accelerates workload execution when reused across queries. However, since workload execution with different candidates is costly, approximate acceleration estimates are used. %the accuracy of which poses additional problem for candidate selection. 
    The accuracy of \textbf{benefit estimation} is essential for the optimality of candidate selection.
\end{mytakeaway}

\subsection{Constraints}

\centeredItalic{What types of constraints are typically considered in candidate selection?}

\begin{center}
\begin{figure}
    \centering
    \includegraphics[width=0.8\linewidth,keepaspectratio]{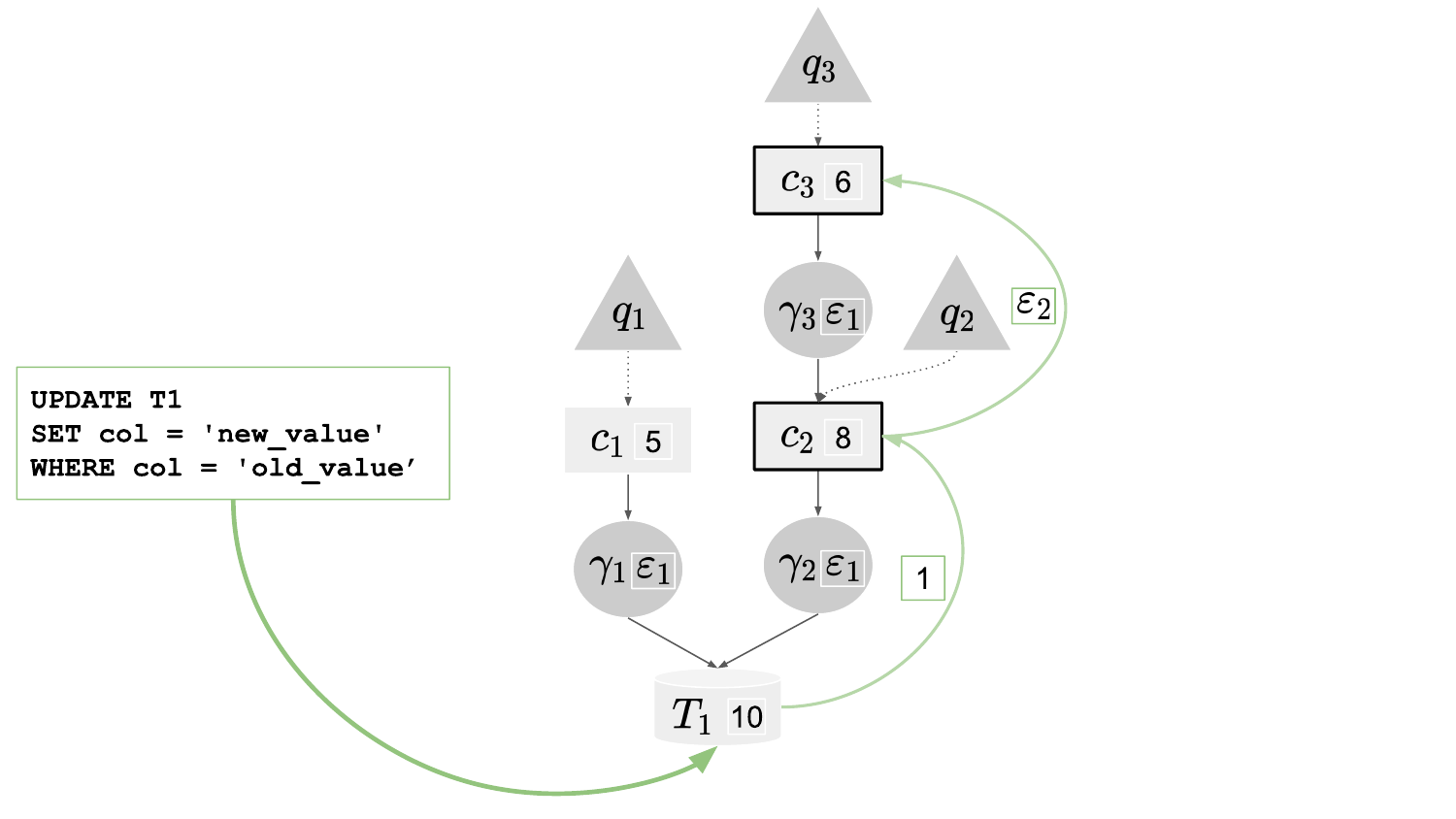}
    \caption{When table $T_1$ is updated, the selected candidates \textit{must} also be updated. And updating candidate $c_3$ \textit{can be accelerated by reusing the updated} common computation $c_2$, which shows that the expense function may have a complex behaviour. The selected candidates are shown in black rectangles and update operations with the corresponding execution times are shown in green.}
  \label{fig:update}
\end{figure}
\end{center}

\noindent It can be shown that the total benefit does not decrease when we select more candidates, so one potentially obtains the highest benefit when all computations are selected for reuse. Obviously, this is impractical, since selecting a candidate $c$ incurs some \textit{expense} $\e{c}$ which is related, e.g., to the disk space used for storing $c$. Similar to benefit, the expense of selecting a candidate depends on other elements in $C$. For example, in VSP, it is necessary to keep the data in materialized views up-to-date, so $\e{c}$ represents the time required to update a view $c$ \cite{gupta1999selection}. Given that other views may be used for the update, the expense $\e{c}$ depends on other candidates in $C$:

\begin{example}
\label{ex: updating}
    Consider the workload from Figure \ref{fig:update}. If $c_2$ is selected, the time of updating $c_3$ will be shorter, since we can \textit{reuse} the \textit{updated} $c_2$:
    $$
        \underbrace{8 + \varepsilon_2}_{\substack{\text{update } c_3 \\ \text{over } c_2}} \text{ vs } \underbrace{10 + \varepsilon_1}_{\text{calculate } c_2} + \underbrace{1}_{\text{update } c_2} + \underbrace{8 + \varepsilon_2.}_{\substack{\text{update } c_3 \\ \text{over } c_2}}
    $$
\end{example}

The non-linear behavior of expense occurs also in \textit{Plan Caching}, in which the problem is to optimally reuse computed plan trees (\textit{not the data itself}). Frequently occurring subtrees can be stored separately and reused when needed (see Example \ref{ex:plan_cache}). Then the expense of storing a plan  depends on whether \textit{any of its subplans is already in the cache}. This may be often the case, since a subplan of an optimal query plan is the optimal plan for the corresponding subquery. As shown in \cite{Delaney2007-zc}, Plan Caching can improve performance, and the noted behavior of $\e{c}$ can help to enhance it:

\begin{figure}
  \centering
  \includegraphics[width=0.8\linewidth,keepaspectratio]{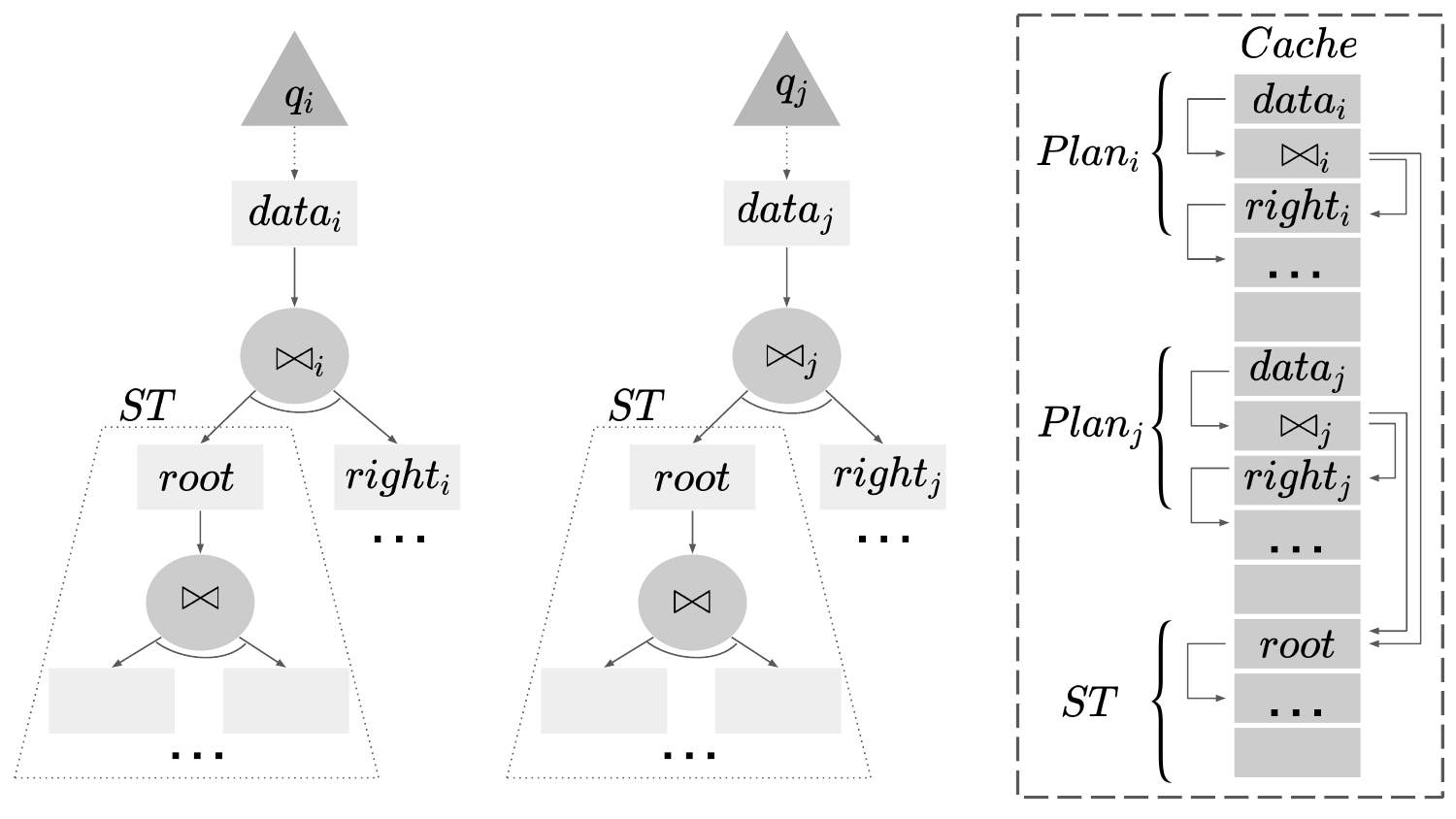}
  \caption{To improve the efficiency of plan caching, it is possible to store common parts of plans in a single instance. This is yet another case of reusing shared computations. On the left: optimal execution plans for queries $q_i$ and $q_j$ with a common subtree \textit{ST}\protect\footnotemark. On the right: plan caching schema, in which the plan for \textit{ST} is stored in memory only once.}
  \label{fig:plan_cache}
\end{figure}
\footnotetext{the set of edges depicted with an arc is and-arc (see Section  \ref{subchapter:query_representation} for more details)}

\begin{example}
\label{ex:plan_cache}
    Let optimal plans for queries $\{q_i\}$ have a common subtree $ST$ (see Figure \ref{fig:plan_cache}). Instead of saving $ST$ for each of plans $\mathcal{P}(\{q_i\})$ we can save it \textit{once}, and refer to it in all its supertrees. Then the saved cache space can be used to speed up $Q$ by caching plans for other queries. We will call this way of caching plans \textit{compressed tree storage}.
\end{example}

In general, one has to deal with the constraint \\$\sum_{c \in C}\e{c} \leq E$, where $E$ is an expense budget, which represents, e.g, the available storage space or time for update. As we will show further in the paper, the problem of the \textbf{non-linearity of expense} makes the selection problem fundamentally more complex. 

\begin{mytakeaway}
    Storing a candidate for reuse comes with an expense. Depending on the type of expense, its behaviour can be highly complex. The \textbf{non-linearity of expense} is one of the main difficulties in solving the selection problem.
\end{mytakeaway}

\subsection{Candidate Selection Problem} 

\centeredItalic{What do all candidate selection problems have in common and what distinguishes them from other Integer Programming problems?}

\noindent From the examples above one can see that in all the types of selection problems considered the nature of candidates is not important: it is essential how the benefits and expenses are computed. Therefore, we now introduce a generalized \textit{Candidate Selection Problem (CSP \footnote{not to confuse with the Constraint Satisfaction Problem}}) which highlights the main aspects of selection tasks. 

Let $\mathbb{C}$ be a set of \textit{candidates} and for $c\in \mathbb{C}$, let $\mathcal{B}_c(\cdot)$ and $\e{c}(\cdot)$ be the benefit and expense for candidate $c$, respectively. The Candidate Selection Problem is to select a subset of candidates $C\subseteq \mathbb{C}$ which gives the maximal total benefit $\mathfrak{B}(C)$ under a given expense budget $E$:
\begin{equation}\label{eq:cs_def}
    \begin{aligned}
    &\mathfrak{B}(C) := \sum_{c \in C} \mathcal{B}_c(C) \to \max_{\{C \subseteq \mathbb{C} | \sum_{c \in C} \e{c}(C) \leq E \}}
    \end{aligned}
\end{equation}

\begin{table*}%[htbp]
    \centering
    \begin{threeparttable}
        \begin{tabular}{cllll}
            \toprule
            \diagbox{\textbf{parameter}}{\textbf{problem}} & View Selection & Index Selection & Query Caching & Plan Caching \\
            \midrule
            candidate space $\mathbb{C}$   & eq-nodes & eq-nodes & eq-nodes & \textbf{subtrees} \\            
            expense $\e{c}(C)$ & non-constant\tnote{a} & non-constant\tnote{a} & constant & non-constant\tnote{b}\\            
            benefit $\mathcal{B}_c(C)$         & \multicolumn{4}{l}{non-constant, computed as the speed-up under reuse of $c$ in plan $\mathcal{P}_C(Q)$} \\
            \bottomrule
        \end{tabular}
        \begin{tablenotes}
            \item[a] under maintenance constraint
            \item[b] when compressed tree storage is used
        \end{tablenotes}
        \caption{All MQO problems can be considered as instances of the Candidate Selection Problem}
        \label{tab:problem_modeling}
    \end{threeparttable}    
\end{table*}

\noindent CSP \textbf{represents all selection problems in Multi-Query Optimization} (see Table \ref{tab:problem_modeling}). It is worth noting that, unlike benefit (which is non-linear in all the selection problems considered), the behavior of expense $\e{c}(\cdot)$ is problem-specific. 

\begin{mytakeaway_main}
    The formulation may seem overly general, but we \textit{deliberately avoid} here specifying a structure of benefits and expenses. As we will see many times in the paper, the common feature in the considered problems is that \textit{objects can be represented as trees, and benefits and expenses can be specified as weights of some paths}. It is this feature of selection problems that opens up various possibilities for optimization. Given the fact that candidates in View Selection, Index Selection, and Plan/Query Caching problems can be also represented by trees, we consider the candidate selection framework to be very promising.
    %the proposed candidate selection framework is very general. 
\end{mytakeaway_main}    

\begin{figure}%[ht]
  \centering
  \includegraphics[width=0.7\linewidth,keepaspectratio]{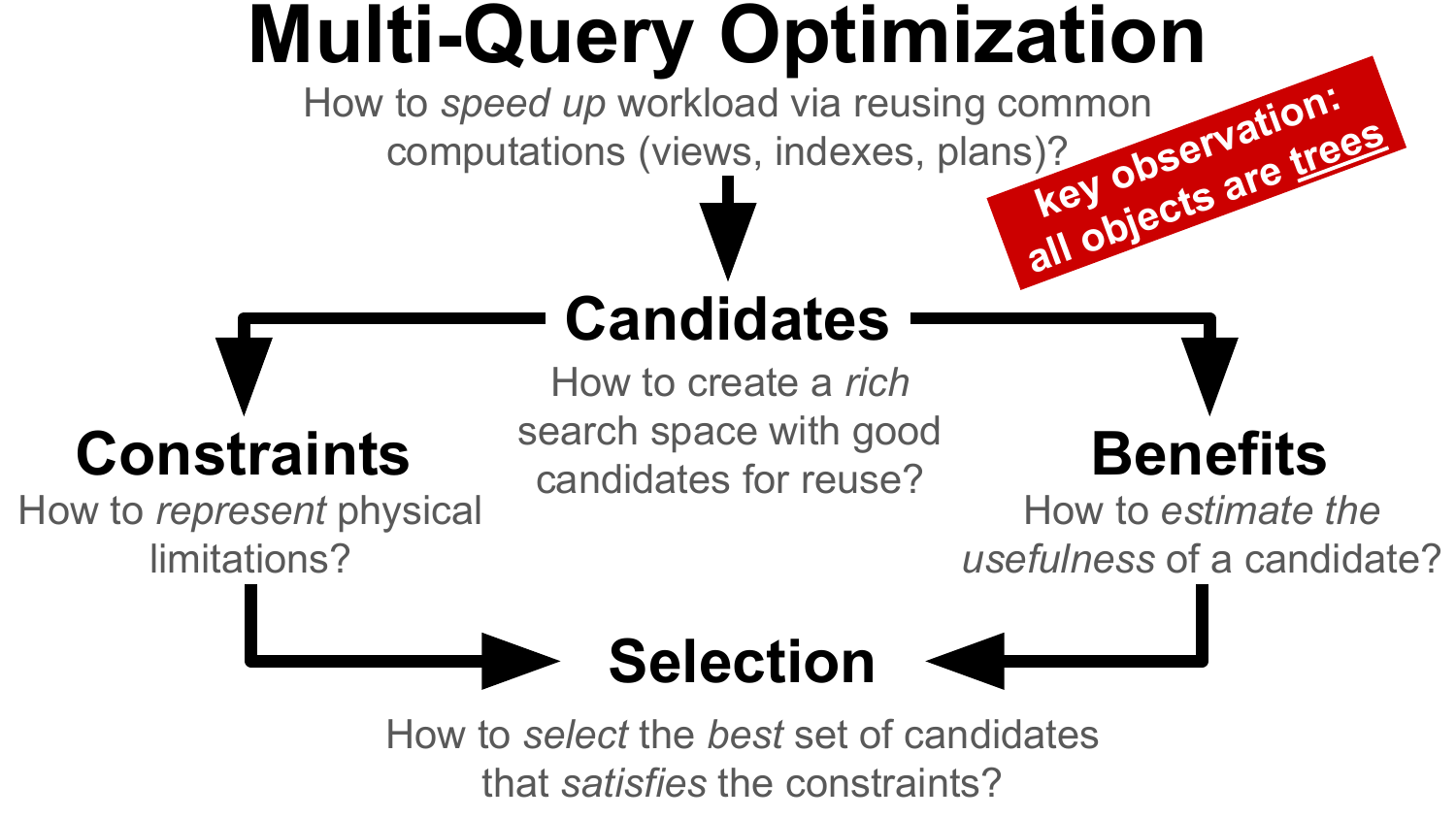}
  \caption{The key observation in our study is that the tree structure of candidates can be used to address virtually \textit{all stages} in solving the View Selection Problem. Since most MQO problems also represent objects as trees, the developed techniques can be successfully reused in these cases as well.} 
  \label{fig:landscape}
\end{figure}

% \begin{mytakeaway}
%     \textbf{Takeaways.}\\
%     All MQO problems can be considered as a special case of the candidate selection problem. The key idea behind this generalization is that \textit{in all MQO instances, candidates can be represented by trees}, which is the \textit{main} property of the candidate structure used in the design of almost all algorithms.
%     %This representation is typically used in the design of selection algorithm
% \end{mytakeaway}

\subsection{Computational Complexity} 
\label{subchapter:theory}

\centeredItalic{What is the complexity of the Candidate Selection Problem? \footnote{we demonstrate the complexity of CSP by giving an overview of the complexity results of its VSP subproblem.} 
Is it possible to quickly find an approximate solution?}

\noindent It is known that under the space constraint VSP in OR-DAG\footnote{a definition is given in Chapter \ref{subchapter:query_representation}} is NP-hard. It has been also proved that a greedy algorithm can give a solution that is at least 63\% of the optimal one (47\%, respectively, for the case when views are selected together with indexes), and this lower bound \textit{can not be improved} in polynomial time \cite{gupta1997selection}. Under the \textit{maintenance constraint} (a limited budget for keeping views up-to-date) a similar result does not hold and solutions obtained greedily \textit{can be arbitrarily bad} \cite{gupta1999selection}. The reason is the non-monotonicity of the benefit per unit space due to the non-linear behavior of $\e{c}
$. In the case of selecting views together with indexes, the accuracy drops from 63\% to 47\% because of the non-monotonicity too, but the reason is the non-linear behavior of the benefit itself. Indeed, we can only benefit from the index if the corresponding view is selected.

In minimizing the total execution time, VSP is known to be \textit{polynomially non-approximable} \cite{karloff1999complexity}. This does not contradict the result with 63\% accuracy of a greedy algorithm, because by using the benefit we implicitly transition to a closely related (i.e., their optimal solutions coincide) but still a different optimization problem. The original optimization objective is the total time $T_C(Q)$, but when using benefits we are maximizing $\mathfrak{B}(C) = T_{\varnothing}(Q) - T_{C}(Q)$. Let $\widehat{C}$ and $C^*$ denote the obtained and the optimal sets of candidates, respectively. Then clearly, from the relationship $\mathfrak{B}(\widehat{C}) \geq k \cdot \mathfrak{B}(C^*)$ the total time in terms of $k$ can not be derived. The reason is that the ratio of $T_{\varnothing}(Q)$ to $T_{C^*}(Q)$ is \textcolor{Maroon}{unknown}:

\begin{equation*}
\Bigg( \frac{\mathfrak{B}(\widehat{C})}{\mathfrak{B}(C^*)} = \frac{T_{\varnothing}(Q) - T_{\widehat{C}}(Q)}{T_{\varnothing}(Q) - T_{C^*}(Q)} \Bigg) = k \Leftrightarrow \frac{T_{\widehat{C}}(Q)}{T_{C^*}(Q)} = k + (1-k) \cdot \textcolor{Maroon}{\bm{\frac{T_{\varnothing}(Q)}{T_{C^*}(Q)}}}
\end{equation*}

For the case when query plans contain non-unary operators (e.g., joins), approximability \textit{is still an open question} for VSP even in the benefit setting under space constraint. Also, to the best of our knowledge, the question of the complexity of VSP in the case of AND-DAG\footnote{definition is given in Chapter \ref{subchapter:query_representation}} \textbf{has also been open} \cite{gupta2005selection}. We \textbf{answer} this question by the following theorem:
%(full proof is provided in Appendix \ref{appendix:np-hardness})

\begin{theorem}
The View Selection Problem over binary AND-DAG under space constraint is NP-hard.
\end{theorem}

\textit{Proof.} The theorem is proved by a reduction of the Knapsack Problem. Suppose we have $n$ items with values $v_i$ and weights $w_{i}$ and our total space budget equals to $W$. We can assume that $\min_i(w_{i})^2 > W$, because otherwise we can ensure this by scaling with an appropriate constant. Consider a join of $n$ tables with independent filtering WHERE-clauses $\sigma_i$ having sizes $w_i + v_i$ and $w_i$ before and after filtering, respectively. Corresponding query could look like:

\begin{lstlisting}[style=sqlstyle]
SELECT * FROM T1 JOIN T2 ... JOIN Tn
WHERE (T1.col1 = const1) AND (T2.col2 = const2)  ... AND (Tn.coln = constn).
\end{lstlisting}

\noindent If we assume that the time of reading and filtering a table equals its size, then view selection (under \textit{any} execution plan of $q$ represented as an AND-DAG) solves the knapsack problem. Indeed, we can only select candidates from filtered tables ($c_i=\sigma_i(T_i)$). Any other candidate in the graph will correspond to the join of at least two filtered tables, i.e. its size will be at least $\min_i(w_{i})^2$, so it will not fit into the space budget. The benefit we get from the reuse of $c_i$ is equal to 
$$
    \underbrace{(w_i + v_i)}_{\text{compute $c_i$ }} - \underbrace{w_i}_{\text{reread $c_i$ }} = v_i,
$$
which is entirely consistent with the selection of the item $i$. Thus the View Selection Problem on AND-DAG of \textit{arbitrary shape} under space constraint is at least as hard as the Knapsack Problem.
\QED

\medskip

\noindent Interestingly, if we consider only plan nodes as candidates (no matter how they are represented), we may miss the optimal solution. Chirkova et al. \cite{chirkova2002formal} showed that, in general, the space of candidates that needs to be considered to find an optimal solution is \textit{infinite}.

To overcome the \textbf{computational complexity} of the selection problem, a plethora of approaches has been proposed including heuristic-based, randomized algorithms, or custom ones which provide solutions close to the optimum only for a fixed pool of queries and specific underlying data. The latter class of algorithms is based on recently proposed Machine Learning approaches. There is also a number of optimizations to reduce the search space and employ the tree structure when computing benefits and expenses. We will discuss these techniques in more detail in the next two sections.

\begin{mytakeaway}
    Even under simple linear constraints, the CSP problem is NP-hard, and the question of polynomial approximability in the case of non-unary operators is still open.
\end{mytakeaway}

%% file: chapters_preparation.tex
\section{Preparation for selection}
\label{chapter:preparation}
We now highlight several important techniques that are used prior to selection to make the overall procedure more efficient.

\subsection{Query Representation}
\label{subchapter:query_representation}

\centeredItalic{How to build a rich candidate space?}

\noindent \textbf{Subexpression Space.}  Although an optimal solution can be missed when \textit{only} subexpressions of queries are used for building a space of candidates (see Example \ref{ex:reuse}), this approach still has several important advantages. First, it makes the process of building the candidate space relatively simple: candidates are searched only among the nodes of the plan obtained from the optimizer and alternative computation paths are omitted, which greatly reduces the search space. Second, since all candidates are subexpressions, their execution statistics can be used to approximate benefits. Alternatively, special techniques are sometimes used to modify subexpressions and build new candidates in the context of a specific workload or MQO instance, which makes the search space richer. We discuss this in more detail at the end of the subsection.

\begin{figure}
  \centering
  \includegraphics[width=0.8\linewidth,keepaspectratio]{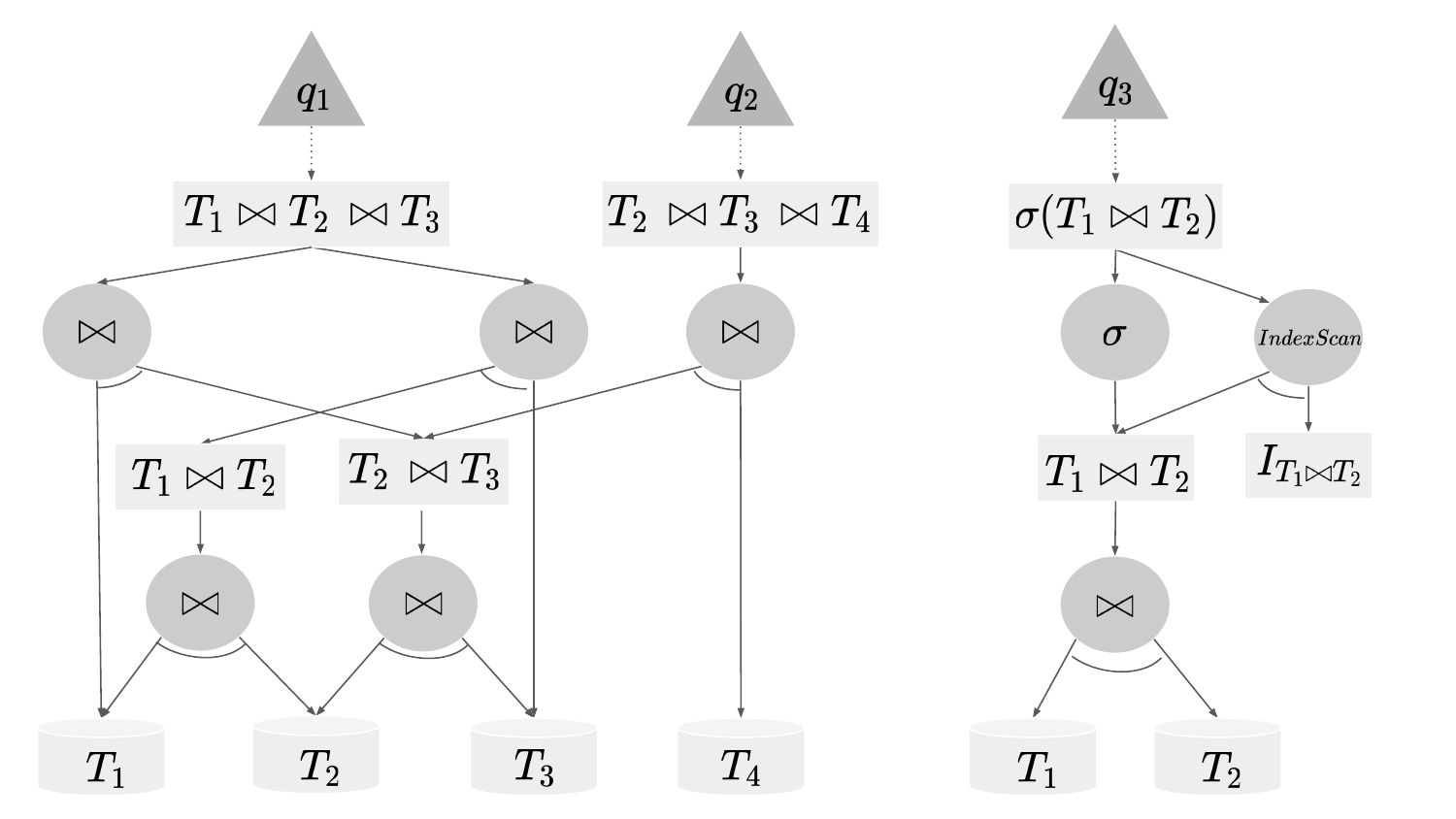}
  \caption{On the left: representation of the workload by using AND-OR-DAG depicting alternative ways of executing query $q_1$, after which the possibility of reusing the common computation $T_2 \bowtie T_3$ to $q_2$ becomes apparent. On the right: representation of alternative execution paths of query $q_3$ by using AND-DAG which reflects the need for both, index $I_{T_1\bowtie T_2}$ and data $T_1 \bowtie T_2$ for \textit{IndexScan} operation.}
  \label{fig:and_dag}
\end{figure}

\noindent \textbf{Representation Frameworks.} 
In case the candidate space is given by subexpressions, there is still a freedom in choosing a representation for expression trees. For example, in order to account for the existence of alternative execution paths, an expression tree can be represented as a \textit{OR-DAG}. This is equivalent to allowing eq-nodes to have multiple children which correspond to different computation paths. To represent non-unary operations (e.g., joins), \textit{AND-DAG} representation framework can be used, which introduces \textit{and-arcs} as several directed edges connected by an arc. This representation takes into account the need to compute \textit{all} operands to execute an operation. Definitions of these classes of DAGs can be found in \cite{gupta2005selection}. This work also introduces a notion of \textit{AND-OR-DAG} which combines the features of the two previous frameworks. 

\begin{example}
    \label{ex:and_dag}
    As we can see in Example \ref{ex:reuse}, the option of using $T_2 \bowtie T_3$ for both queries must be available. This can be implemented by representing multiple computation paths for node $T_1 \bowtie T_2 \bowtie T_3$ with an or-arc (see the left part of Figure \ref{fig:and_dag}). The arcs under joins are and-arcs which indicate the need compute \textit{all} operands for a join operation. The right part of the Figure shows that and-arc can represent a plan which uses an index: to use the index (op-node \textit{IndexScan}) we need both, the data (eq-node $T_1 \bowtie T_2$) and the index itself (eq-node $I_{T_1 \bowtie T_2}$).
\end{example}

In OLAP scenarios, one typically uses aggregations specified by GROUP BY clause over a set of columns. One can consider these aggregation queries as vertices in the hypercube given by the set of table attributes. The 'can-be-computed-by' relation over queries then coincides with the inclusion relation over the sets in the hypercube, so this representation is called \textit{Hypercube Lattice} (or \textit{Data Cube}) in the literature. If a workload $Q$ is represented in the form of a Data Cube, there is \textit{no need to build an expression forest}, since all query relationships are \textit{already given} by the cube. However the Data Cube introduces a space of $2^n$ vertices (where $n$ is the number of table attributes) in which all possible aggregate queries can be represented. If aggregation is ranked according to some granularity (e.g., by day, week, or year), one also has to consider the \textit{granularity hierarchy}. For example, the \textit{Product Graph} (\textit{direct product} in \cite{harinarayan1996implementing}) framework provides this feature. 

Choosing a representation framework is an important step, because the complexity of selection strongly depends on the way query plans are represented (see Section \ref{subchapter:theory}).

\noindent \textbf{Computation Sharing.} 
As we have already noted, the set of nodes of an expression forest may not contain an optimal candidate. To deal with this problem, one can use additional techniques to enrich the search space. For example, one can employ the technique of additional computations:

\begin{example}
\label{ex:compensation}
    Consider a workload of two queries $\sigma_A (T_1)$ and $\sigma_B (T_1)$ given in Figure \ref{fig:reuse}. Suppose, we have a space budget $E = 9$, which is insufficient to store the answers to both queries ($E = 9 < 6 + 6 = \e{c_1} + \e{c_2}$). Then we can generate and store an \textit{additional computation} $c_3 = \sigma_{A \lor B} (T_1)$ which can be shared between the two queries. This gives the maximal possible benefit 
    $$
        \mathcal{B}({c_3}) \approx 2 \cdot (\underbrace{10}_{\text{read } T_1} - \underbrace{7}_{\text{read } c_3}) = 6
    $$ (compared to the cases when $c_1$ or $c_2$ are stored), but implies the need of \textit{extra} operation, (called as \textit{compensations} in \cite{cochraneapplying}) for fetching the required data. For example, to answer query $q_1$, we must apply extra selection $\sigma_A$ to candidate $c_3$, which, in turn, is also obtained by using selection $\sigma_{A \lor B}$. Note that, individually for each query, the $\sigma_{A \lor B}$ filter incurs computational overhead, but it makes the candidate more reusable and accelerates the entire workload.
\end{example}

Also various `rewriting' strategies based on relational properties \cite{yang1997framework} and identities \cite{roy2000efficient, roy1998practical} can be used. By choosing rewrite rules and the order of their application to the nodes of individual query plans one can influence the number of common nodes among plans, as well as their sizes. This may have a positive impact on the quality of the resulting solution. We provide more details on this technique in Section \ref{subsubchapter:heuristics_space_design} where we describe the corresponding algorithms.

\begin{figure}
  \centering
  \includegraphics[width=0.8\linewidth,keepaspectratio]{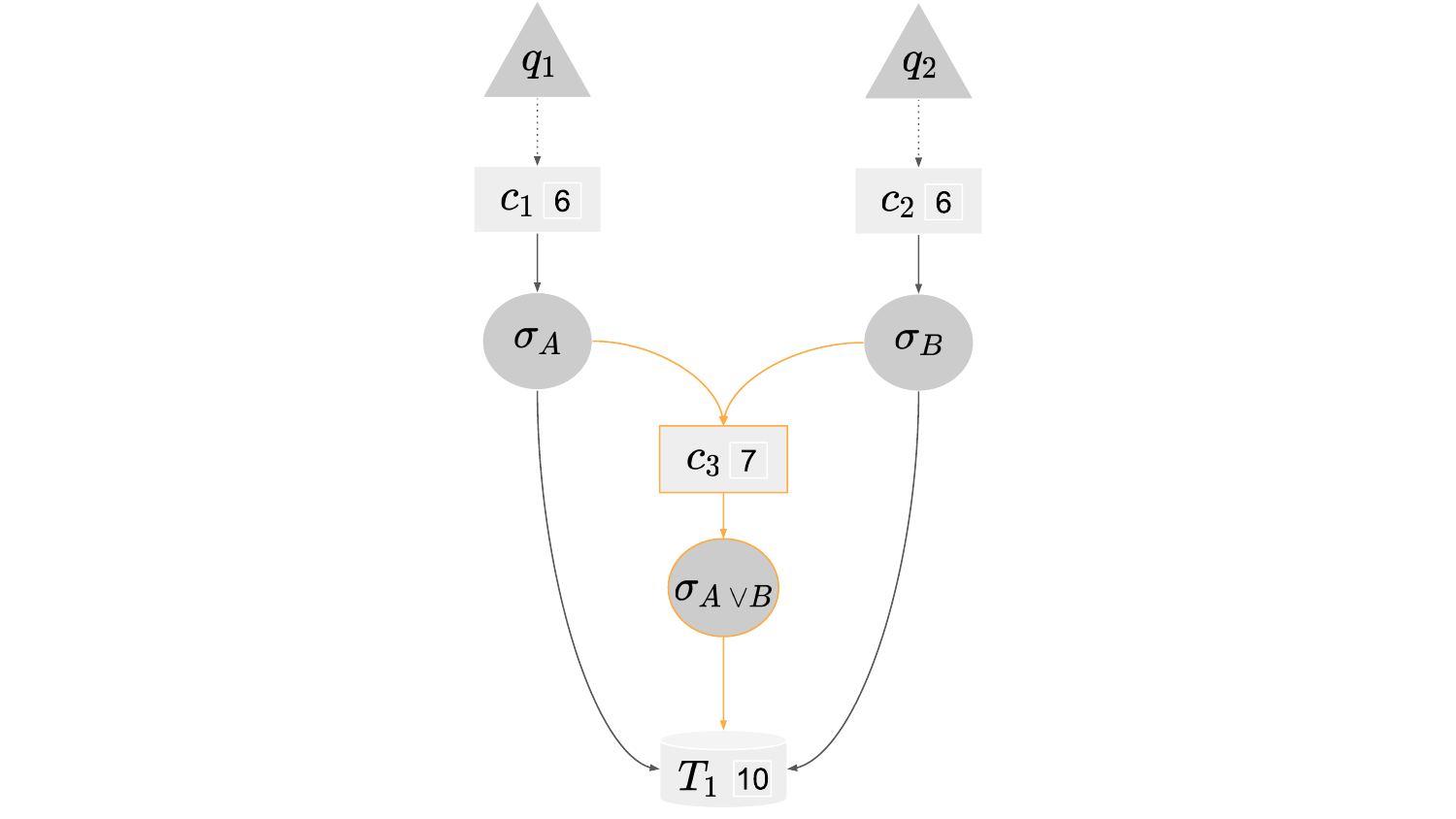}
  \caption{Introducing $\sigma_{A \lor B}$ enables computation reuse. Despite the potential overhead of creating and maintaining $\sigma_{A \lor B}$, one can achieve an overall speed up of the workload. The additional computation and the way of its reuse are shown in orange.}
  \label{fig:compensation}
\end{figure}

\begin{mytakeaway}
    Subexpression space is a good starting point in building a search space for candidates. If an appropriate candidate is not in the search space, one can expand it by using compensation techniques or by rebuilding query plans to take into account the context of a given workload. To consider several alternative ways to execute a query one can integrate plans within a representation framework with OR arcs.
\end{mytakeaway}

\subsection{Benefit Estimation}
\label{subchapter:benefit_estimation}

\centeredItalic{What are the approaches to benefit computation?}

\noindent \textbf{Lightest Path and Cost Models.} 
In terms of the expression tree framework, query latency is determined by the weight of the lightest computation path from the eq-node corresponding to the query to the leaves\footnote{or saved eq-nodes, because we can read them instead of executing from scratch} of the tree. To compute the weight of a computation path we need to take into account arc types in the following manner: 1) if an or-arc is encountered, we can continue the computation path through \textit{any} of the children, 2) for a and-arc, we have to compute \textit{all} the children.
A \textit{useful property of tree-calculated benefits} is that the time of executing a node $c$ is affected only by the selection of its descendants. But it is important to note, that the benefit of a candidate $c$ \textit{still depends on its parents} too, since their selection may change the lightest computation path preventing $c$ from being used (see Example \ref{ex:benefit}). In Section \ref{subchapter:heuristics}, we discuss how this observation can be employed to speed up algorithms.

To compute the weight of a computation path, one also needs to know the estimates for operation latencies and data sizes. These are typically obtained from the optimizer and thus, may be inaccurate. %As the optimizer may make mistakes, the estimates obtained this way may be inaccurate. This effect amplifies in situations where a node $c$ is used multiple times by several queries of workload. 
To combat this, predictive models have been proposed, which, employ, e.g., run-time statistics on the execution of queries for making predictions about the future \cite{jindal2018computation}.

\noindent \textbf{End2End Modeling.} 
Instead of reducing the problem to the search for the lightest computation path, one can use the optimizer in \textit{what-if} mode \cite{zilio2004recommending}. The idea is to simulate the situation when a candidate computation of interest is reused. Then the candidate benefit is the difference between the values with and without the candidate predicted by the optimizer. Since the optimizer will know that the selected candidates are available, it may return a plan that was not previously considered as optimal. That is, it will already be able \textit{to take into account the presence of workload} when making plans. However the nature of benefits is highly complex and we have to compute it by a call to the optimizer each time the set of candidates $C$ changes, which is highly inefficient. Alternatively, one can train a ML model to answer questions like \say{what is the benefit $\mathcal{B}_c$ of a candidate $c$ for a given workload $Q$ if we have already selected candidates $C$?} \cite{yuan2020automatic}.

\noindent \textbf{Integer Programming.} 
There are also approaches in which the problem of node selection and reuse is reduced to an Integer Programming problem \cite{yang1997algorithms}. This is a simple way to obtain an exact solution for the selection problem (provided accurate estimates are available), from which it is also easy to compute the benefit $\mathcal{B}_c$ of each candidate $c$. Indeed, it can be obtained as the difference between the execution time of $c$ and the time of its reuse, multiplied by the number of uses of $c$ in the optimal solution. We note however that computing an exact solution to an IP problem to obtain benefits makes little sense for the following reasons. First, if an optimal solution is found, then then benefits are not needed anymore. Second, in order to formulate the selection problem as an IP, one needs to introduce a large number of additional variables, which would make search for a solution prohibitively long. Instead, one can first fix a set of candidates $C$, and then solve simpler problems of finding the optimal way to use them \textit{independently}, for each query $q$ from a workload. These subtasks can be solved in parallel giving an approximation of benefits. In Section \ref{subsubchapter:hybrid_modern_approaches} we discuss several ways to implement this idea.

\begin{mytakeaway}
Benefit can be computed by traversing the expression forest (where the weights are derived from estimated costs of the corresponding operations) or by using predictions made by the optimizer in `what-if' mode. The latter method allows for taking the entire workload into account. To enhance accuracy, the candidate space can be extended and operation costs can be estimated by identifying similar queries in historical data or by training a separate predictive model.
\end{mytakeaway}

\subsection{Dealing with Constraints}
\label{subchapter:constraints}

\centeredItalic{How are contraints modeled?}

\textbf{Specialized Solutions}. 
\noindent Recall that the constraint in CSP is given as $\sum_{c \in C} \e{c}(C) \leq E$ and in some cases the expense function $\e{c}(\cdot)$ may be non-linear. That is, for $C = C_1 \sqcup C_2$ it holds in general that
$$
    \sum_{c \in C}\e{c}(C) \neq \sum_{c_1 \in C_1}\e{c_1}(C_1) + \sum_{c_2 \in C_2}\e{c_2}(C_2).
$$
\noindent This poses challenges to selection algorithms: for example, a greedy algorithm can no longer guarantee any \% of accuracy, in contrast to the case of space constraints. In part, this can be overcome by specialized solutions. For example, in \cite{gupta2005selection} Gupta et al. made a theoretical analysis of the behavior of the greedy algorithm and proposed to use special  \textit{inverted set tree} structures to achieve the desired accuracy of 63\%. However, the complexity of this solution is exponential in general.

\noindent \textbf{Penalization.} 
Even simple constraints with constant $\e{c}$ can pose difficulties. For example, in randomized and genetic algorithms, the set of candidates is built iteratively, and at each step the action to be taken for a candidate $c$ must depend on its benefit $\mathcal{B}_c$. The problem here is two-fold. First, the benefit takes into account neither the expense of the candidate, nor the remaining budget. Second, situations, in which a constraint is violated, must be handled accordingly. Simply avoiding such situations may result in a poor strategy. Indeed, the path which goes only through admissible solutions from the current solution to a good local optimum may either not exist or be very long. To address this, a penalization approach has been proposed \cite{lee2001speeding}, which tries to account for the presence of a constraint in the form of some penalty (\textit{regularizer}). The simplest way to implement this is the \textit{substract mode}. The idea is to use the penalty function $\phi(C) = \max(0, \sum_c \e{c}(C) - E)$ and compute the benefit as $\mathfrak{B}'(C) = \mathfrak{B}(C) - r \cdot \phi(C)$, where $r$ is some regularization coefficient. The value of $r$ affects \say{how much we do not want to violate the condition $\sum_{c \in C} \e{c}(C) \leq E$}. Important is to choose the right coefficient $r$, because if it is chosen badly, the algorithm will tend to \say{non-violation of constraints} instead of optimization. This can be avoided by measuring everything in the same units. For example, in VSP, if we know the rates of computational resources at runtime, as well as the price per unit of disk space, we can express everything in money and maximize the total profit \cite{yuan2020automatic}. This makes sense, because there is no problem with disk space and memory availability nowadays, the only issue is payback. 

\noindent \textbf{Stochastic Solutions}. 
If a selection algorithm uses benefits only to compare candidates (as in the local search algorithms), we can use \textit{stochastic ranking} instead of penalization  \cite{lee2001speeding}. In short, the idea is to employ a comparison that favors more cheap candidates with probability $1-p$ with no regard to their benefits \cite{runarsson2000stochastic}.

\begin{mytakeaway}
    In scenarios when the expense for a candidate does not depend on the selected candidates, it suffices to satisfy the space constraint. Otherwise, one also has to consider a maintenance type constraint, which may require a modification of the search algorithm. It is useful to model constraints as penalties to simplify the search landscape or take them into account with some probability like in stochastic ranking.
\end{mytakeaway}

%% file: chapters_algorithms.tex
\section{Selection Algorithms}
\label{chapter:algorithms}
Now we analyze in detail the techniques sketched in the previous sections, we provide a modern classification of the selection algorithms employing these techniques, and discuss the main trends in the development of such algorithms. Our exposition is primarily based on an analysis of algorithms for View Selection, with the emphasis on the techniques that can be reused for solving various instances of the Candidate Selection Problem in Multi-Query Optimization.

\input{subchapters_exhaustive}

\input{subchapters_heuristic}

\input{subchapters_randomized}

\input{subchapters_hybrid}

\input{subchapters_algo_summary}

%% file: subchapters_exhaustive.tex
\subsection{Exhaustive Search}
\label{subchapter:exhaustive}

\centeredItalic{How large is the search space in general?}

\noindent Assume a workload $Q$ consists of queries that involve joins over different subsets of tables $\{T_j\}_{j=1}^n$. Then, in the worst case, the size of the search space (the number of candidates among which it makes sense to look for a solution) is $2 ^ n$ (the number of all possible joins). As the benefit $\mathcal{B}_c$ of an individual candidate non-linearly depends on the set of selected candidates $C$, a naive search for an optimal selection would have to enumerate all possible subsets of $C$, which is already of double exponential size $2^{2^n}$. This kind of algorithm was proposed in \cite{ross1996materialized}, but its complexity is prohibitive for the size of modern databases.

%% file: subchapters_heuristic.tex
\subsection{Heuristics}
\label{subchapter:heuristics}
\subsubsection{Optimizations of Greedy Algorithm.}
\label{subsubchapter:heuristics_selection}

\centeredItalic{How can a greedy algorithm take into account the complex interaction of benefits and constraints? How does the structure of the search tree help to speed up computations?}

\noindent In many scenarios, it suffices to compute a non-optimal, but good enough solution within a constrained time budget. To implement this, one can search for a solution only in a certain subexpression space. Plenty of heuristics has been proposed which aim at selecting the most useful candidates \cite{jindal2018selecting}: 

\begin{enumerate}
    \item \textbf{Topk-freq}: selecting the most common candidates
    \item \textbf{Topk-utility}: selecting candidates with the maximal benefit they can provide for an individual query
    \item \textbf{Topk-TotalUtility}: selecting candidates with the highest total benefit for entire workload
    \item \textbf{Topk-NormTotalUtility}: selecting candidates with the highest specific total benefit for entire workload per unit of space
\end{enumerate}

\noindent These heuristics do not take the non-linearity of benefit into account and thus, they can sometimes provide poor solutions.

The paper \cite{harinarayan1996implementing} addresses this shortcoming. The authors propose a greedy algorithm in which, at every step, the currently selected set of  candidates $C$ is expanded with a candidate $c$, which gives the largest total benefit $\mathfrak{B}(C \cup c)$. It was proved in \cite{karloff1999complexity} that this algorithm guarantees the accuracy of at least $(1-\frac{1}{e}) = 63\%$, and no polynomial time algorithm can do better. We  note however that the constraint considered in \cite{harinarayan1996implementing} is the number of selected candidates, not the space they occupy.  

Gupta et. al in \cite{gupta1997selection} adopted a similar idea for the case of the space constraint. They proposed to measure the benefit change, when a candidate $c$ is added, per unit of space it occupies:
$$
    b_c(C) := \frac{\mathfrak{B}(C\cup c) - \mathfrak{B}(C)}{\mathcal{E}(C\cup c) - \mathcal{E}(C)},
$$
The authors proposed an algorithm for OR-DAG and AND-DAG frameworks which in both cases provided the accuracy guarantee of 63\%. For the general AND-OR-DAG framework, the authors proposed a \textit{AO-Greedy} algorithm which is in fact a greedy algorithm over a specially defined \textit{intersection graph} structure. The size of this structure is exponential in general, so the proposed algorithm is no longer polynomial in the size of the expression forest.

For the setting when indexes are selected together with views, Gupta et al. proposed an \textit{inner-level greedy algorithm}. The option of selecting indexes breaks the monotonicity property, because one can not use an index if the corresponding data is not selected (see Example \ref{ex:and_dag}), and as a consequence, the benefit of this index is zero. The inner-level greedy algorithm guarantees a 47\% accuracy in OR/AND-DAG frameworks and it is based roughly on the following idea. At each iteration, the algorithm first searches for a view which has the highest benefit along with its best indexes. Then the set of candidates $C$ is extended either by this view and its indexes, or by one new index (probably, for another view) having the highest benefit. This procedure allows one to avoid the following  problem: if a view is useful \textit{only} with an index, then it \textit{can not be selected} by a basic greedy algorithm. Indeed, the index will not be selected, because without the view its benefit is 0, and the view is not is selected because without an index its benefit insufficient. To adapt the method to the AND-OR-DAG framework, the authors proposed a generalization in the form of a \textit{r-level greedy} algorithm, which is able to guarantee a certain \% of accuracy in situations with more complex dependencies in benefit. 

For the setting with the maintenance constraint in the OR-DAG framework, the authors proposed to use an \textit{inverted-tree set} structure to regain the monotonicity property. An estimate of 63\% accuracy for a greedy algorithm using this set structure is guaranteed, but the number of these sets is exponential in the worst case. 

In \cite{mistry2001materialized}, two optimizations of the greedy approach were proposed by taking into account the tree structure in benefit computation. Since at each step the greedy algorithm searches for the best candidate $c$ to expand $C$ with, one needs to frequently evaluate the total benefit of sets that differ just in a single element. Since the benefit of the nodes is often computed by tree traversal, one can \textit{cache benefits for nodes} and recompute them only when the choice of a new candidate $c$ has an impact on them. Along with this technique, the authors proposed a \textit{coarse heuristic} that assumes that the node benefit can only decrease\footnote{this is not true in general, since there is also a non-linear update time $l_c$ component in the overall execution time, which may decrease as new candidates are added}. With this heuristic, if the nodes are stored in the descending order of the (previously computed) benefits, then iteration over all candidate nodes outside of $C$ can often be avoided in finding the best next candidate to expand $C$.

\subsubsection{Reduction of Candidate Space}
\label{subsubchapter:heuristics_space_reduction}

\centeredItalic{What heuristics can be used to narrow down the search space? How does the tree structure help to prune the search space?}

\noindent Clearly, the running time of a selection algorithm depends not only on the selection procedure itself, but also on the size of the search space. Several approaches to reduce the search space have been proposed in the literature. In \cite{baril2003selection}, each individual query plan is traversed in the \textit{level-order}, and only vertices from the level that give the highest benefit are taken into the candidate space $\mathbb{C}$. This allows one to get rid of the complex dependencies in benefits and expenses as it is guaranteed that the parents and children of the candidates will not be selected.

In \cite{agrawal2000automated}, the authors suggested to consider only those candidates which refer to the \say{most interesting} tables. A table $T_1$ is considered to be more interesting than a table $T_2$ if the total execution time of queries touching $T_1$ is higher than that of queries touching $T_2$. Based on the selected set of tables, the authors defined the candidate space according to the following idea: \say{\textit{if a candidate is not in the optimal plan of a query from a given workload then it is unlikely to be useful}}. For each query $q$ from a  workload they define a candidate space $\mathbb{C}_{q}$ as the union of all subsets of the most interesting tables touched by $q$ (the authors also analyzed conditions in queries and formulated conditions for grouping or selection over joins of interesting tables). Then by using a \textit{greedy(m, k)} algorithm\footnote{greedy(m, k) algorithm first searches exhaustively for a  subset of \textit{good} $k$ elements from the total set of size $m$ and then expands this set greedily} they select from them the most interesting candidates $C_{q}$ for each query $q$. The candidate space $\mathbb{C}$ for the selection problem is defined as the union of these sets. 

Gupta et al. \cite{gupta2005selection} proposed a different approach based on the following idea: instead of reducing the entire search space apriory, we can avoid exploring those parts of the space where there is definitely no better solution. To implement this, the authors adapted the ideas of $A^*$ algorithm \cite{nilsson1982principles}. They considered the situation when all eq-nodes of the expression forest must be computed ($\mathbb{C} = Q$). Their $A^*$-like algorithm is as follows. The search space is represented as a \textit{search tree}, with vertices $v = \langle C, \mathbb{C}_{seen}\rangle$ representing information about visited candidates $\mathbb{C}_{seen} \subseteq \mathbb{C}$ and selected ones $C \subseteq \mathbb{C}_{seen}$. There is an edge $v \rightarrow v'$ if the vertex $v'$ is obtained from $v$ by adding a candidate $c$, i.e., $v' = \langle C \sqcup \{c\}, \mathbb{C}_{seen} \sqcup \{c\}\rangle$ or $v' = \langle C, \mathbb{C}_{seen} \sqcup \{c\}\rangle$. The goal is to reach a vertex $v^*=\langle C^*, \mathbb{C}\rangle$ in which the  selected set of candidates $C^*$ gives the minimal execution time for the given workload. 

The algorithm tries to find this vertex by exploring the search tree \textit{only in the most promising directions}. To do this, it assigns to each vertex $v=\langle C, \mathbb{C}_{seen}\rangle$ an estimate of the minimum time of executing the entire workload $\mathbb{C}$, provided we saved only $C$ for the execution of $\mathbb{C}_{seen}$. The principal problem is to get an accurate estimation, because it directly influences the extent to which the search space is reduced. This problem can be solved as follows. Assume the optimal way to expand the current set of selected candidates $C$ is $C' \subseteq \mathbb{C} \setminus \mathbb{C}_{seen}$. Then, by taking into account the additivity of the execution time, the best execution time for the workload achievable from the current vertex $v$ can be decomposed as follows:
$$
    T_{C \sqcup C'}(\mathbb{C}) =  T_{C \sqcup C'}(\mathbb{C}_{seen}) + T_{C \sqcup C'}(\mathbb{C} \setminus \mathbb{C}_{seen}).
$$ 
The key point of the algorithm is the traversal of vertices $c \in \mathbb{C}$ in a \textit{topological order}, which guarantees that the execution time of queries from $\mathbb{C}_{seen}$ remains unchanged under future extensions of $C$, i.e. $T_{C \sqcup C'}(\mathbb{C}_{seen}) = T_C(\mathbb{C}_{seen})$ (the reason for this is explained in Section \ref{subchapter:benefit_estimation}). Then instead of estimating the value $T_{C \sqcup C'}(\mathbb{C})$ we can compute the lower estimate $\widehat{h}(C)$ \textit{of a simpler value} which is the execution time of the rest of the workload $\mathbb{C}\setminus\mathbb{C}_{seen}$. To do this the authors employ a separate greedy algorithm (see \cite{gupta2005selection} for details). At each iteration, the $A^*$-like algorithm selects a vertex for which the predicted execution time $\widehat{h}(C) + T_{C}(\mathbb{C}_{seen})$ is minimal. This significantly reduces the search space in practice. The algorithm returns an exact solution, but in the worst case it has to explore the entire graph of size exponential in $\mathbb{C}$.

Labio et al. \cite{labio1997physical} employed similar ideas in a slightly different scenario. They considered a setting in which \textit{already materialized} views $Q$ need to be updated efficiently. One way to achieve this is to spend some resources on \textit{materialization of new computations} which speed-up updates of $Q$. The problem can be formulated in terms of the Candidate Selection Problem as follows. For a given workload $Q$ (queries that describe already materialized views), select a set of candidates $C$ from an appropriate candidate space $\mathbb{C}$ such that a) the total update time for $Q$ with $C$ is minimal and b) $Q\subseteq C$. The latter requirement is important, since the views $Q$ are already materialized and we have to spend resources on updating them. The authors adapted $A^*$ algorithm for this scenario. In this setting, for vertices $\langle C, \mathbb{C}_{seen} \rangle$, it is required to estimate the minimum time of the total update after expanding $C$ with a set $C'$ provided $Q \subseteq (C' \sqcup C)$. The total update cost \textit{(UC)} can obviously be divided into two parts:
$$
    UC_{C' \sqcup C}(C \sqcup C') = UC_{C' \sqcup C}(C) + UC_{C' \sqcup C}(C').
$$ 
\noindent If the vertices are traversed in the topological order  then the dependence of $UC_{C' \sqcup C}(C)$ on $C'$ can be avoided. The remaining cost $UC_{C' \sqcup C}(C')$ was estimated from below, based on the extreme situation when all unconsidered vertices are updated in the most expensive way (taking into account the selected candidates $C$) and are used when updating other data whenever possible. In practice, with this approach the authors were able to obtain a reduction of the search space by 4 orders of magnitude.

\subsubsection{Design of Candidate Space.}
\label{subsubchapter:heuristics_space_design}

\centeredItalic{What heuristics can be used to enrich the search space?\\ And how does the tree structure of candidates allow for efficiently computing benefits?}

\noindent Heuristic approaches are also used for a targeted design of a candidate space which should contain a good solution. For example, in \cite{yang1997framework} a method to build an expression forest $\mathcal{F}$ takes into account the fact that individually optimal plans may not contain the most useful computations for executing a given workload $Q$. First, an optimal plan $\mathcal{P}({\{q\}})$ is built for every query $q\in Q$. Then the selection and projection operations are pushed up in expression trees, which makes the candidates larger and hence, implies potentially more savings from their re-use. Next, the resulting plans $\mathcal{P}'({\{q\}})$ are merged into a forest $\mathcal{F}$ in a cost-based manner. For robustness, the authors even built several such forests at once. 
Finally, all the selection and projection operations in $\mathcal{F}$ are pushed down\footnote{which is a standard technique for reducing the size of processed data}, and then a greedy algorithm is run. The best set of candidates found among all forests is returned as a solution.

In addition to the concept of interesting tables (which we discussed in the previous subsection), in \cite{agrawal2000automated} the authors also introduced a \textit{MergeViewPair} algorithm to take into account the nature of MQO. The algorithm iteratively merges pairs of candidates $c_1, c_2$ into a single candidate $c$, while a) preserving the possibility of using $c$ instead of $c_1$ and $c_2$, and b) guaranteeing a small computational overhead of using $c$ instead of $c_1$ and $c_2$. It is worth noting that using $c$ is often slightly more expensive, since it contains data from (at least) both $c_1$ and $c_2$. But the number of situations in which $c$ can be reused is much larger, which is important for optimization of the entire workload (Example \ref{ex:reuse} illustrates this technique). At the final stage, a \textit{greedy(m, k)} selection algorithm is applied to the candidate set $\mathbb{C}$ obtained by merging.

The question of building an expression forest $\mathcal{F}$ was also raised in \cite{roy2000efficient}. To obtain forests that would contain a good solution, they proposed two algorithms. In the first one, \textit{Volcano-SH (SHared)}, individually optimal plans are first built independently and then the techniques of \textit{computation sharing} described in Section \ref{subchapter:query_representation} are applied. In the second one, \textit{Volcano-RU (ReUse)}, plans are built sequentially. This makes possible to figure out which nodes belong to optimal plans for other queries. The execution time for these nodes is deliberately underestimated, so that they are more often found in optimal plans for other queries. Because of this the resulting expression tree has more common nodes and provides options for their reuse. Node benefits are computed in a special way. To calculate the benefit of reusing a candidate $c$ one needs to know a) the time of computing $c$ and b) the frequency $num\_uses(c)$ of using $c$. Therefore, the benefit depends on both the children and parents of a candidate node. The first dependency is eliminated by traversing vertices in the topological order and the second one by heuristically estimating $num\_uses(c)$ by the number of  ancestors of $c$: $num\_uses(c)\geq \#ancestors(c)$ \footnote{this holds, because the framework considered does not reflect alternative computation paths, so there is no way to avoid computing a child when executing an ancestor}. This approach allows for quickly and accurately estimating the benefit.

%% file: subchapters_randomized.tex
\subsection{Randomized Algorithms}
\label{subchapter:randomized_algo}

As we can see, all known heuristic methods are either rather time consuming ($A^*$ algorithm) or they return an approximate solution (by a greedy algorithm). We now consider randomized algorithms which provide an efficient alternative to heuristic ones.

\subsubsection{Random Sampling.} 
\centeredItalic{What can be done in the case of an extremely small budget to select good candidates?}

\label{subsubchapter:random_sampling}
\noindent In \cite{kalnis2002view}, several randomized algorithms for solving the View Selection Problem were proposed. The authors used the Data Cube framework, in which the search space is represented by bit strings of length equal to the number of nodes in the cube. Each bit indicates whether the corresponding candidate is selected. One of the simplest algorithms considered in \cite{kalnis2002view} is \textit{Random Sampling} which looks as follows: a) randomly sample a bit string b) check whether all constraints are satisfied, and c) estimate the benefit. The item that satisfies the constraints and has the highest benefit is returned as the answer. The algorithm can be a solution of choice in scenarios when the computation budget is very limited \cite{10.5555/645920.672975}.

\subsubsection{Local Search.} 
\label{subsubchapter:local_search}

\centeredItalic{How to define candidate neighborhood in order to adapt classical local search algorithms?}

\noindent In the same paper the authors proposed more involved \textit{Iterative Improvement} (abbreviated as II) and \textit{Simulated Annealing} (SA) algorithms. To explain the main ideas behind them we need to define the notion of neighborhood of two solutions. Let us introduce three types of actions: 1) select new candidate, 2) replace one candidate with another one, 3) remove candidate. Two solutions are called neighbors if one of them can be obtained from another by a single action. II algorithm implements random transitions only to neighbors $C'$ with a higher benefit $\mathfrak{B}(C')$. In  SA algorithm, a (random) transition to neighbors with a lower benefit is possible, but the probability of such a transition is the less the smaller the benefit of the neighbor is. The intuition behind this is as follows: local optima may well be connected by a short path via candidates with a smaller benefit, but frequent transitions to less useful solutions make the search longer. As II algorithm makes transitions only to useful neighbors, it converges to good solutions rather quickly, although it may get stuck in isolated regions. In turn, the quality of solutions obtained by SA is higher. To combine the advantages of both algorithms, a \textit{Two-Phase Optimization procedure (2PO)} is  employed, which was previously proposed in \cite{ioannidis1990randomized}. By using II algorithm, 2PO first converges to an area of potential interest and then explores it in a more detail by SA algorithm.

In \cite{derakhshan2006simulated}, an algorithm based on Simulated Annealing is proposed. In this work, solutions are considered to be neighbors if their representation strings differ only at one position. The algorithm is rather simply adapted to a parallel computing scenario \cite{derakhshan2008parallel}, in which communication between processes is not required when moving to neighbors: several SA procedures are run independently and the best solution they find is returned as the final answer.

\subsubsection{Genetic Algorithm.} 
\label{subsubchapter:genetic_algo}

\centeredItalic{How to define the concepts of mutation and crossover?}

The genetic algorithm is a well-known approach to solve NP-hard problems; it is based on a search procedure inspired by the principles of natural selection and genetics. The search is carried out by an iterative improvement of generations, each of which is represented by a population. From each generation, with the help of \textit{mutations} and \textit{crossovers}, a new generation is created, from which the best individuals are \textit{selected}. The principle step here is to represent candidates as a population. To both a) reflect the existence of multiple individual query plans and b) represent the entire set of candidates, Horng et al. \cite{horng2003applying} proposed the following schema. By using a special \textit{query-plan string} ($qps$), they record which execution plans for queries are selected, and concatenate it with a \textit{view string} ($vs$), which represents selected candidates. Mutation is implemented by changing the value in a random position in the string, while the crossover is implemented as a \textit{cut-and-swap} operation separately on $vs$ and $qps$. To keep the population size limited, a random number of the best candidates is kept according to their benefits. Also, at each iteration, the candidates are improved via local search, i.e., in fact, the proposed procedure refers to the class of \textit{Genetic Local search algorithms} \cite{kolen1994genetic}.

In the above presented schema, as in most evolutionary algorithms, candidates are compared based on their benefit. But if the optimization problem includes a space constraint then it must also be taken into account. Indeed, if one candidate has a slightly less benefit than another one, but implies a significantly less expense, then it might be worth selecting it. In \cite{lee2001speeding}, the authors suggested to use the technique of penalty functions, which allows for a) choosing cheaper candidates from those having the same benefit and b) skip solutions that violate constraints. This approach helps to \textit{reduce the length} of the shortest path to the optimum and to \textit{avoid complex search landscapes} where valid regions are separated by invalid ones. 

An alternative approach is randomized \textit{stochastic ranking}, in which candidates are compared with probability $p$ by their benefit, and with probability $1-p$ by the size of the remaining expense budget. In \cite{yu2003materialized} Yu et al. proposed an evolutionary algorithm which generalizes the selection step. Population selection is implemented similar to the bubble sort, where stochastic comparison is used instead of the standard comparison. Sorting ends if no permutations occurred at an iteration. After this, the first $k$ individuals are selected as the result.

%% file: subchapters_hybrid.tex
\subsection{Hybrid algorithms}
\label{subchapter:hybrid}

One more important step in the development of selection algorithms was made by hybrid approaches which compensate for weak points of some classes of methods with the advantages of the others. 
\subsubsection{Early Approaches.} 
\label{subsubchapter:hybrid_early_approaches}

\centeredItalic{How to compensate for the inaccuracy of fast heuristic algorithms by using randomized ones?}

\noindent One of the first ideas in this direction was proposed in \cite{zhang2001evolutionary}. The authors considered a setting in which a query has several plans, and proposed a two-stage algorithm. The first stage is to select a plan for every query and build an expression forest $\mathcal{F}$ over them. The second stage is to select candidates from $\mathcal{F}$. At each of these stages, two basic algorithms are used: a greedy \cite{gupta1997selection,yang1997framework} and an evolutionary \cite{horng2003applying} one. The authors tried 4 combinations of algorithms in total. In experiments they came to the conclusion that going beyond individually optimal plans for queries can greatly improve the quality of obtained solutions. Also, the most accurate solution was obtained when the problem at stage 2 was solved by an evolutionary algorithm. This confirms the hypothesis that \textit{correctly estimated benefits are essential for making good selection}, because evolutionary algorithms do not rely on greedy heuristics and provide more accurate benefit estimates.

In \cite{zilio2004recommending}, the authors combine the ideas of greedy and randomized approaches for efficient selection of views together with indexes. At the first step, similarly to the techniques described in Section \ref{subsubchapter:heuristics_space_design}, they expand the candidate space by finding common subexpressions and compensations (see Example \ref{ex:compensation}). Benefits are calculated by using the optimizer's \textit{'what-if' mode} (see Section \ref{subchapter:benefit_estimation}). Based on the obtained values, a greedy algorithm is applied which has the property that in case an index is selected at some iteration, the algorithm tries to select the corresponding view simultaneously (provided there is enough space). Then an iterative improvement is performed for the solution obtained by the greedy algorithm, which randomly picks several unselected candidates instead of selected ones. This allows for compensating the non-optimality of the greedy algorithm.

\subsubsection{Modern Approaches.}
\label{subsubchapter:hybrid_modern_approaches}

\centeredItalic{How can a problem be decomposed to support efficient parallelization?\\ How can Machine Learning help to improve components of the selection algorithms?}

\noindent In \cite{jindal2018computation}, the View Selection Problem is considered in a practical DBaaS scenario. The authors note that \% of common computations between queries from real workloads is typically quite large, while savings from their reuse can be as high as 40\% (of the available computing resources). By using the recurrent nature of queries in their scenario, the authors manage to cope with two problems at one shot. First, to deal with inaccurate cost estimates from the optimizer the authors propose to use statistics on previously executed queries. This is known in the literature as the \textit{feedback loop} technique. Secondly, they implement view selection in an online scenario (i.e., when the workload is not known in advance) by using information about the workload observed in the past and by assuming that queries in the future will be similar. To quickly search for relevant statistics and check for useful materialized views, they employ a so-called  \textit{signature technique} that takes into account the \textit{recurrent} nature of queries and is used as an efficient search filter. The authors also take into account that queries are executed in parallel and therefore, decisions on the materialization of views are also made in parallel. As a consequence, the same view can be created multiple times, which wastes both memory and time. To deal with this, the authors propose a special \textit{early materialization} technique. In short, the idea is to: a) not to materialize the same computation multiple times, b) make the materialized computation available for reuse until the end of the execution of the query that triggered its materialization. Even if the triggering query is rolled back, the materialized view remains available for reuse. 

In \cite{jindal2018selecting}, the same authors proposed new ideas on the selection procedure in the form of the \textbf{BigSubs} algorithm. In their approach, selection is initially formulated as a ILP problem. However, in order to model the non-linear behavior of functions, one has to introduce a large number of variables, which makes the resulting ILP problem infeasible. The main idea of BigSubs is to use an \textit{iterative} algorithm which allows for splitting the original problem into a big number of independent subtasks. The decision which candidates should be selected at each iteration is made by a \textit{random} $flip$ function. After a set of candidates is fixed, the optimal ways to use them are decided \textit{independently} for each of the queries by using solutions to ILP subtasks. Based on the estimated benefit of selected candidates the distribution of the $flip$ function is changed and then the next iteration is performed. To optimize the solution of ILP subtasks, a number of heuristics is proposed. One part of them is essentially based on the fact that candidates are represented as trees, and the other part takes into account the nature of the interaction of queries in the workload. By using the idea of problem decomposition and by employing a vertex-centric graph processing model (like Giraph \cite{malewicz2010pregel} or GraphLab in \cite{low2014graphlab}), the authors managed to cope with the size of the original problem, which was beyond the capacity of the reference ILP solver Gurobi \cite{gurobi}.

The state-of-the-art solution to VSP is given by \textbf{RLView} algorithm from \cite{yuan2020automatic} which extends the BigSubs approach by replacing the procedures of random flipping and benefit estimation with Machine Learning algorithms. This work is not limited to the online scenario and assumes that there is enough budget to factorize workload subexpressions into semantic equivalence classes (e.g., by using \textit{Equitas} equivalence checker \cite{zhou2019automated}), which provide a fairly rich candidate space. Then in each equivalence class, the cheapest representative is chosen, which significantly reduces the set of candidates. The authors approach the problem of inaccurate predictions of the optimizer by using a \textit{Wide-Deep model}, which is a neural network capable of modeling both linear and highly non-linear dependencies of the execution time on input parameters \cite{cheng2016wide}. The model is trained on  collected statistics, which is similar to the feedback loop technique \cite{stillger2001leo, marcus2019neo}. Various ways of encoding strings, keywords, and table schemas in queries are employed, over which recurrent LSTM networks \cite{greff2016lstm} are run to capture the overall structure of the query. The constraints in the selection problem are taken into account in the form of a regularizer. The regularization coefficient is defined by converting everything into a universal measure unit (money). 

The intuition behind the RLView approach can be explained as follows. In fact, \say{\textit{BigSubs has no memorization ability and it does not converge to a global optimal solution, because there is no information sharing between different iterations}}. Therefore, inspired by the success of Reinforcement Learning \cite{sutton2018reinforcement}, the authors essentially proposed a randomized (\textit{but informed}) search procedure based on a Markov Decision Process (MDP). Typically, a MDP is defined by 4 parameters: a state space $S$, an action space $A$, a  distribution $P_a(s, s')$ of the probability of transition between states $s$ and $s'$ by an action $a$, and a distribution $R_a(s, s')$ of reward $r$ when moving from state $s$ to $s'$ by $a$. The task is to find a strategy $\pi: S \times A \rightarrow \{0, 1\}$ which gives, on average, the highest total reward. The fundamental difference from BigSubs is that this model of the flip function a) takes into account the current state of $s$ and b) learns over time. 

In RLView algorithm, a state $s$ is a set of selected candidates and a description of how to use them, an action $a$ means a change in the decision on the selection of a candidate, and a reward $r$ reflects a change in the total benefit due to an action. The optimal view selection strategy is searched by using Q-Learning \cite{watkins1992q}. In particular, the authors build a neural network \textit{DQN} \cite{mnih2013playing, wang2016dueling}, which is trained  to predict the potential benefit from each of the actions in a given state. In experiments, the authors established 3 SotA results on different datasets and also confirmed the hypothesis that \textit{the accuracy of estimated benefits plays one of the key roles} in solving the selection problem. Both RLView and BigSubs demonstrated best results when using a predictive model instead of optimizer estimates.

%% file: subchapters_algo_summary.tex
\subsection{Summary on Algorithms}
\label{subchapter:algo_summary}

\centeredItalic{What are the common points for building efficient selection algorithms?}

\noindent As we have shown, the algorithms for solving the selection problem can be roughly divided into three classes.  Heuristic algorithms, based on intuitive assumptions for estimating the benefit of candidates and for reducing the candidate space, aim at providing fast approximate solutions. Randomized algorithms try to improve over heuristic ones by offering a trade-off between the accuracy and speed of computations, but they provide no guarantees on the quality of obtained solutions. Recent advances have been made with hybrid solutions that utilize the strengths of approaches from different domains, including Machine Learning. 

We observe that there is a number of common points in these algorithms and we summarize the most important ones below. 

\noindent \textbf{The choice of candidate space}. 
The quality of obtained solutions strongly depends on the space of candidates. To obtain a good candidate space one can use plan building strategies, plan  merging techniques, and methods for finding common subexpressions (Sections \ref{subsubchapter:heuristics_space_design} and \ref{subsubchapter:hybrid_early_approaches}). It is also plausible to use space reduction methods, which provide more flexibility in building an efficient selection algorithm (Section \ref{subsubchapter:heuristics_space_reduction}).

\noindent \textbf{Tree structure of candidates.} 
In selection problems, candidates are typically tree nodes while benefits are given as weights of the lightest computation paths. Relationships between candidates and the way benefits / expenses are computed can greatly reduce the search space and speed up computations (Section \ref{subsubchapter:heuristics_space_reduction}). 

\noindent \textbf{Dependence on accurate cost estimates}.
Several implementations have confirmed the intuitively clear fact that \textit{the quality of solutions obtained by selection algorithms strongly depends on the accuracy of benefit estimates for candidates} (Sections \ref{subsubchapter:hybrid_early_approaches} and \ref{subsubchapter:hybrid_modern_approaches}). For example, recent advances in solving the View Selection Problem are due to approaches which \textit{do not rely on estimations from DB optimizer}. Modern algorithms employ history-based (feedback loop) techniques and ML based prediction models. 

\noindent \textbf{Modularity of solutions and the use of ML.} 
There is a general trend of developing modular solutions which find a compromise between advantages and disadvantages of several types of algorithms (Sections \ref{subsubchapter:local_search} and \ref{subsubchapter:hybrid_modern_approaches}). Since the dependencies between objects involved in the Candidate Selection Problem are highly complex, it is hard to model them explicitly. However, a large amount of data describing these dependencies is available. Given the fact that ML is an excellent tool for discovering complex patterns in data, it is natural to believe that \textit{ML will be used extensively for the development of selection algorithms}.

\begin{mytakeaway}
\textit{What can be taken as a baseline solution to the selection problem?} One can use a greedy algorithm with unit benefits \cite{gupta1997selection}. For example, to take into account the presence of indexes when solving the View Selection Problem, it is recommended to try a modification of the greedy algorithm and inner-level algorithm \cite{gupta1997selection}.
\end{mytakeaway}

\begin{mytakeaway}
\textit{How to get rid of full search in general?} One can try a general greedy(m,k) scheme \cite{agrawal2000automated} or use the technique of truncating the search space by means of a special graph traversal and heuristics \cite{gupta2005selection}.
\end{mytakeaway}

\begin{mytakeaway}
\textit{What if my algorithm still takes a long time to run?} Keep only `interesting' candidates \cite{agrawal2000automated}, or perform a cheap filtering, ignoring complex dependencies between them via level-order traversal \cite{baril2003selection}. In case of an extremely small optimization budget, it is recommended to apply random sampling techniques \cite{kalnis2002view}.
\end{mytakeaway}

\begin{mytakeaway}
\textit{What if solution accuracy is insufficient, but exponential search is unaffordable?} Then in addition to expanding the candidate space and improving the accuracy of the estimates, it is recommended to use randomized algorithms \cite{kalnis2002view}. 
\end{mytakeaway}

\begin{mytakeaway}
\textit{How to balance between time and accuracy in general?} One can try to decompose the problem into several levels, each of which can be solved by a separate algorithm \cite{zhang2001evolutionary,zilio2004recommending}. Also the iterative scheme can be used, which allows one to consider query optimization independently \cite{jindal2018computation,jindal2018selecting,yuan2020automatic}. The number of iterations in this approach can be controlled, and subtasks can be parallelized.
\end{mytakeaway}

%% file: chapters_applying.tex
\section{Application}
\label{chapter:applying}
%\centeredItalic{How can the analysis and systematization of the techniques be useful in practice?}

%We have already given some tips and also shown what difficulties arise in Candidate Selection problems, what techniques exist to solve them and how they are combined to create efficient algorithms. 
In this section, we show how the analysis presented in this paper can be used \textit{to improve the accuracy} and \textit{to significantly speed up} the state-of-the-art RLView and BigSubs algorithms for view selection.

\subsection{SotA solution schema.} 
Recall that BigSubs and RLView are iterative algorithms, in which every iteration consists of two steps. First, a $flip$ function is called to select candidates for the current iteration. Then, the algorithm determines how to optimally utilize the selected candidates. The results of this step are used to refine the behavior of the flip function. One iteration is shown in Algorithm \ref{alg:iterative_algo}, where $u_{ij}$ represents the benefit of candidate $c_j$ for query $q_i$. %(i.e., $\mathcal{B}_{\{c_j\}}(\{c_j\})$ for workload $Q = \{q_i\}$). 
The vector \textbf{stats} contains information about gained benefits, expenses, and on how candidates are currently reused. The variable $z_j$ is a \textit{boolean} one representing the selection decision for candidate $c_j$, $e_j$ is the expense of storing it, and $E$ is the total available budget.

\begin{algorithm}[H]
\caption{One iteration of the BigSubs / RLView algorithm}\label{alg:iterative_algo}
\begin{algorithmic}
\Require $Q$, $\{u_{ij}\}$, $E$, $\{e_j\}$, $\{z_j\}$, \textbf{stats}
\Ensure Updated \textbf{stats} and selection decisions $\{z_j\}$

\For{$c_j \in \mathbb{C}$}
        \State $z_j = flip\_function(\textbf{stats}, E, \{e_j\})$ \Comment{Should we select candidate $c_j$?}
\EndFor 

\For{$q_i$ in $Q$}
    \State $\textbf{stats}.update(ILP\_feedback(q_i, \{z_j\}, \{u_{ij}\}))$ \Comment{How useful is each selected candidate for query $q_i$?}
\EndFor 
\end{algorithmic}
\end{algorithm}

The primary difference between RLView and BigSubs is that the flip function in RLView is modeled using Reinforcement Learning, while in BigSubs, it is a parameterized randomized procedure. Both solutions reduce the problem of optimally utilizing a candidate (selected in the first step) to ILP. Since the candidates are \textit{already selected}, the problem can be considered \textit{independently} for each query. This reduces the dimensionality of the problem and enables parallel computations. However, modeling the problem by ILP requires to introduce many additional variables and constraints.

To assign a utility to each query-candidate pair, the authors define $u_{ij}$ as the benefit $\mathcal{B}_{c_j}({c_j})$ for workload ${q_i}$, which is independent of other selected candidates and equals $\mathcal{B}_{{c_j}}' \cdot n\_uses_{c_j}(\textcolor{Maroon}{\boldsymbol{\{c_j\}}})$. This lead to the following linear optimization objective:
\[
\sum_{i: q_i \in Q}\sum_{j: c_j \in \mathbb{C}} u_{ij} \cdot y_{ij}, \quad \forall i,j:\,\, y_{ij} \leq z_j,
\]
where $y_{ij}$ is a binary variable indicating whether candidate $c_j$ is used by query $q_i$, and the constraint $\forall i,j:\,\, y_{ij} \leq z_j$ ensures consistency of selection, i.e., unselected candidates can not be used.

To ensure that that all $n\_uses$ are consistent and this objective accurately reflects the true benefit $\mathfrak{B}(C)$, the authors imposed a constraint preventing the simultaneous use of candidates $c_j$ and $c_{j'}$ if the subtree corresponding to one of them is contained within the subtree of the other. This constraint avoids overestimating the benefit when candidates with nested subtrees are selected together ($y_{ij} = y_{ij'} = 1$) and excludes conflict of their $n\_uses$. For example, consider the expression tree from Example \ref{ex:benefit}. Suppose we reuse candidates $c_3$ and $c_4$ to speed up query $q_3$. From the optimization perspective, it would be most beneficial to use both $c_3$ and $c_4$ which gives the following benefit:
\[
\underbrace{[2+\varepsilon]}_{\mathcal{B}_{c_3}(\{c_3\})} + 
\underbrace{[4+2\varepsilon]}_{\mathcal{B}_{c_4}(\{c_4\})}.
\]
However, this is \textit{impossible} in practice. If candidate $c_4$ is used to execute query $q_3$, its subtree $c_3$ is no longer required, and thus the benefit from reusing $c_3$ cannot be received. To model this constraint, the authors introduced numerous additional variables to account for candidate nesting, which significantly increased the dimensionality of the problem and slowed down search for a solution. However, as we demonstrate next, this constraint is too strict:

\begin{example}
\label{ex:strict_constraint}
%The restriction that prohibits the simultaneous reuse of candidates \( c_j \) and \( c_{j'} \), when one of their corresponding subtrees in the expression forest is nested within the other, is generally an unnecessarily strict constraint that prevents achieving the optimal solution. The only needed constraint is the prohibition of their reuse within the \textit{same} computation flow.
%\end{example}
%\begin{proof}
Consider the expression tree shown in Figure \ref{fig:strict_constraints}, for which the corresponding query can be conceptually represented as

\begin{lstlisting}[style=sqlstyle]
WITH FILTERED_NODES AS (SELECT * FROM GRAPH_TABLE WHERE <some node condition>)
SELECT ...
FROM FILTERED_NODES AS FROM_NODE
JOIN FILTERED_NODES AS TO_NODE
    ON <some graph condition between FROM_NODE and TO_NODE>
JOIN NODE_ATTR
    ON <join condition between FROM_NODE and NODE_ATTR>.
\end{lstlisting}

\noindent The query plan consists of \textit{two} computation flows, in which candidate $c_1$ can be reused. Notably, despite the nesting of candidates $c_1$ and $c_2$, both candidates can be reused \textit{simultaneously} if they are processed in \textit{different} computation flows. Let us verify that this is the most optimal candidate reuse strategy under the constraint that only 30 units of space are available for storing candidates. To begin, recall that the execution cost of a candidate is determined by the execution cost of its operation and all its child-operands:

\[
ex\_cost(node) = ex\_cost(node.operation) + \sum_{child\text{-}operand} ex\_cost(child\text{-}operand).
\]

\noindent For simplicity, we assume the following base costs for simple operations

\[
reuse\_cost(c) = |c|, \quad ex\_cost(scan(T)) = ex\_cost(filter(T)) = |T|, \quad ex\_cost(T_1 \bowtie T_2) = |T_1| \cdot |T_2|.
\]

\noindent Additionally, we assume a selectivity of 0.5 for all operations, including filters and joins. Based on these assumptions and initial sizes, the execution costs for the candidates are as follows:

\[
\begin{aligned}
ex\_cost(c_1) &= ex\_cost(scan(T_1)) + ex\_cost(filter(T_1)) = 20 + 20 = 40, \\
ex\_cost(c_2) &= ex\_cost(c_1) + ex\_cost(scan(T_2)) + ex\_cost(c_1 \bowtie T_2) = 40 + 4 + 10 \cdot 4 = 84, \\
ex\_cost(c_3) &= ex\_cost(c_1) + ex\_cost(c_2) + ex\_cost(c_1 \bowtie c_2) = 40 + 84 + 10 \cdot 20 = 324.
\end{aligned}
\]

\noindent From this, we can immediately calculate the benefits of a single reuse of each individual candidate:

\[
\mathcal{B}_{c_1}' = 40 - 10 = 30, \quad
\mathcal{B}_{c_2}' = 84 - 20 = 64, \quad
\mathcal{B}_{c_3}' = 324 - 100 = 224.
\]

\noindent Clearly, when only candidate $c_1$ is reused, the total benefit is $\mathcal{B}_{c_1}' \cdot \boldsymbol{2} = 60$. Reusing candidate $c_2$ alone yields a benefit of $\mathcal{B}_{c_2}' \cdot 1 = 64$. However, if both candidates are reused simultaneously, considering the computation flow constraints, the total benefit $\mathfrak{B}(\{c_1, c_2\})$ is $\mathcal{B}_{c_1}' \cdot \underline{1} + \mathcal{B}_{c_2}' \cdot 1 = 30 + 64 = 94$. Since candidate $c_3$ cannot be reused due to insufficient  storage budget, we conclude that the restriction on simultaneously reusing nested queries is \textit{unnecessarily strict} in this case and prevents from reaching the optimum.
\end{example}

\begin{mytakeaway}
    It is only important to prohibit the reuse of candidates like in the example above within the same computation flow. Modeling such a problem in ILP terms looks intractable. In the following section, we provide a way to adopt this more flexible constraint, which preserves solutions with \textit{optimal} total benefits, while reducing the \textit{exponential} ILP algorithm to a \textit{polynomial-time} graph-based algorithm.
\end{mytakeaway}
%To take into account this, one would need to forgo binary variables $y_{ij}$ and instead consider the utilization of candidate $c_j$ by query $q_i$ separately for each computational flow. 

\subsection{Our Optimization.} 
Note that at the update step of Algorithm \ref{alg:iterative_algo} there are actually \underline{no constraints}: all the candidates that can be used to speed up the workload \textit{are already selected}. For this reason, finding the fastest way to use them can be made more \textit{efficient} and it \textit{does not have to impose any constraints} on the joint choice of conflicting candidates.

\begin{theorem}
The second step in Algorithm \ref{alg:iterative_algo} is computable in polynomial time while admitting simultaneous reuse of nested candidates.
\end{theorem}
\textit{Proof.} Since a reuse conflict can arise only when some node and its child are reused within the same computational flow, a single traversal in the topological order can be used to decide for each node of the expression forest whether this node \textit{or} some candidates from its subtree should be reused. Indeed, when processing a particular node ($\texttt{node}$), we already know the maximum benefit ($\texttt{max\_subtree\_benefit}$) achieved by selecting some of the children from its subtree, and we can compare it with the benefit of reusing the node itself ($\texttt{node.reuse\_benefit}$). A pseudo-code of this procedure is given in Algorithm \ref{algo:compute_subtree_benefits}.

\begin{algorithm}
\caption{Compute Maximum Benefit for all Subtrees}
\label{algo:compute_subtree_benefits}
\begin{algorithmic}[1]
\State \textbf{Input:} Tree structure with nodes as candidates
\State \textbf{Output:} \texttt{max\_subtree\_benefit}: Maximum benefit for each subtree
\Procedure{MaxBenefitFromSubtree}{\texttt{node}}
    \State \texttt{total\_child\_benefit} $\gets 0$
    \ForAll{\texttt{child} $\in$ \texttt{node.children}}
        \State \Call{MaxBenefitFromSubtree}{\texttt{child}}
        \State \texttt{total\_child\_benefit} $\gets$ \texttt{total\_child\_benefit} $+$ \texttt{max\_subtree\_benefit}[\texttt{child}]
    \EndFor
    \State \texttt{max\_subtree\_benefit}[\texttt{node}] $\gets \max($\texttt{total\_child\_benefit}, \texttt{node.reuse\_benefit}$)$
\EndProcedure
\State \Call{MaxBenefitFromSubtree}{\texttt{root}} \Comment{Compute \texttt{max\_subtree\_benefit} for all subtrees in \textbf{one} traversal}
\end{algorithmic}
\end{algorithm}

After we decide which selected candidates should be reused to achieve the maximum total benefit, it is necessary to calculate the number of reuse times for each of them. To compute this, we traverse the expression forest in the reverse topological order, counting the number of incoming computational flows. As soon as a 'selected-for-reuse' node, i.e., a node satisfying the condition
\[
\texttt{max\_subtree\_benefit(node) = node.reuse\_benefit},
\]
is encountered during the traversal, the subtree traversal must be stopped to avoid early conflicts.

\begin{figure}
\centering
  \includegraphics[width=0.8\linewidth,keepaspectratio]{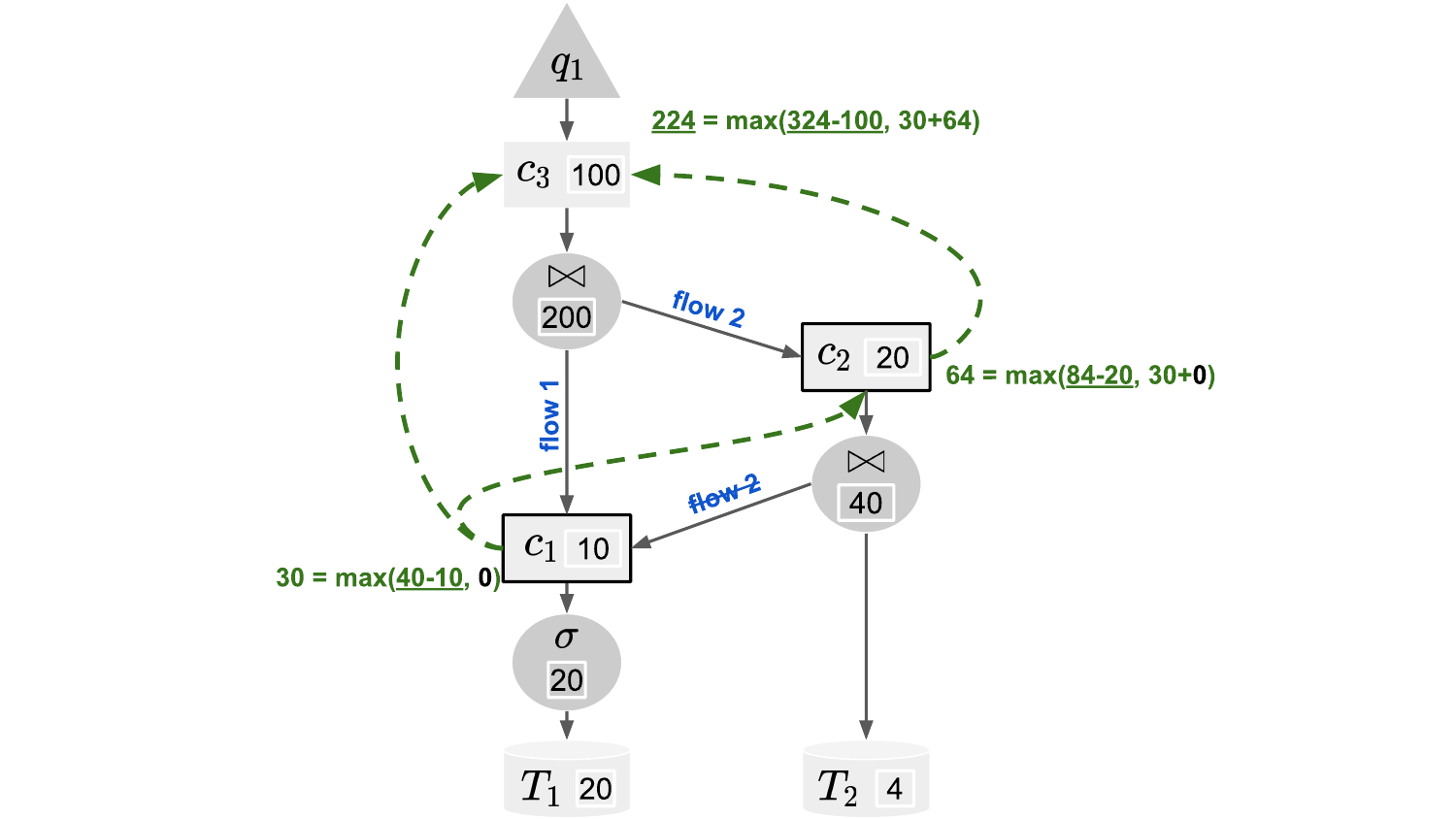}
  \caption{Prohibiting the simultaneous reuse of candidates \( c_j \) and \( c_j' \), when one of their corresponding subtrees in the expression forest is nested in the other, is a too hard constraint which prevents finding the optimal solution. A more relaxed constraint should only prohibit their reuse within the \textit{same} computation flow. Here, both candidates \( c_1 \) and \( c_2 \) should be reused simultaneously (but in different computational flows) to achieve optimal performance. Green arcs depict the process of computing the maximum possible benefit derived from subtrees, while blue arcs illustrate the traversal of computational flows to count the number of reuses for 'selected-for-reuse' candidates.}
\label{fig:strict_constraints}
\end{figure}

By the definition of the procedure above
we obtain an optimally possible reuse of the selected candidates. This can be also illustrated by using Example \ref{ex:strict_constraint}. Indeed, in the first step, we calculate the maximum possible benefit for each subtree in linear time (see the green arrows in Fig. \ref{fig:strict_constraints}). As a result of this, we mark candidates \( c_1 \) and \( c_2 \) as 'selected-for-reuse'. After this, during the top-down tree traversal, each of the two computational flows will independently reach \( c_1 \) and \( c_2 \), respectively, marking them for reuse (see the blue arrows in Fig. \ref{fig:strict_constraints}). As we have shown, this is the optimal way to reuse the candidates.

Thus, by using the proposed method to find the best way of reusing selected candidates and by using Equation \eqref{eq:benefit_decomposition} for benefit decomposition (Sect. \ref{chapter:preliminaries}), we can compute the \textbf{stats} vector of values in \textit{linear} time from Algorithm \ref{alg:iterative_algo}, as opposed to the exponential time required in the worst case when using an ILP solver. Note that we did not impose strict constraint on the reuse of nested candidates and the proposed method provides an optimal solution. \QED

\begin{figure}
  \centering
  \includegraphics[width=0.6\linewidth,keepaspectratio]{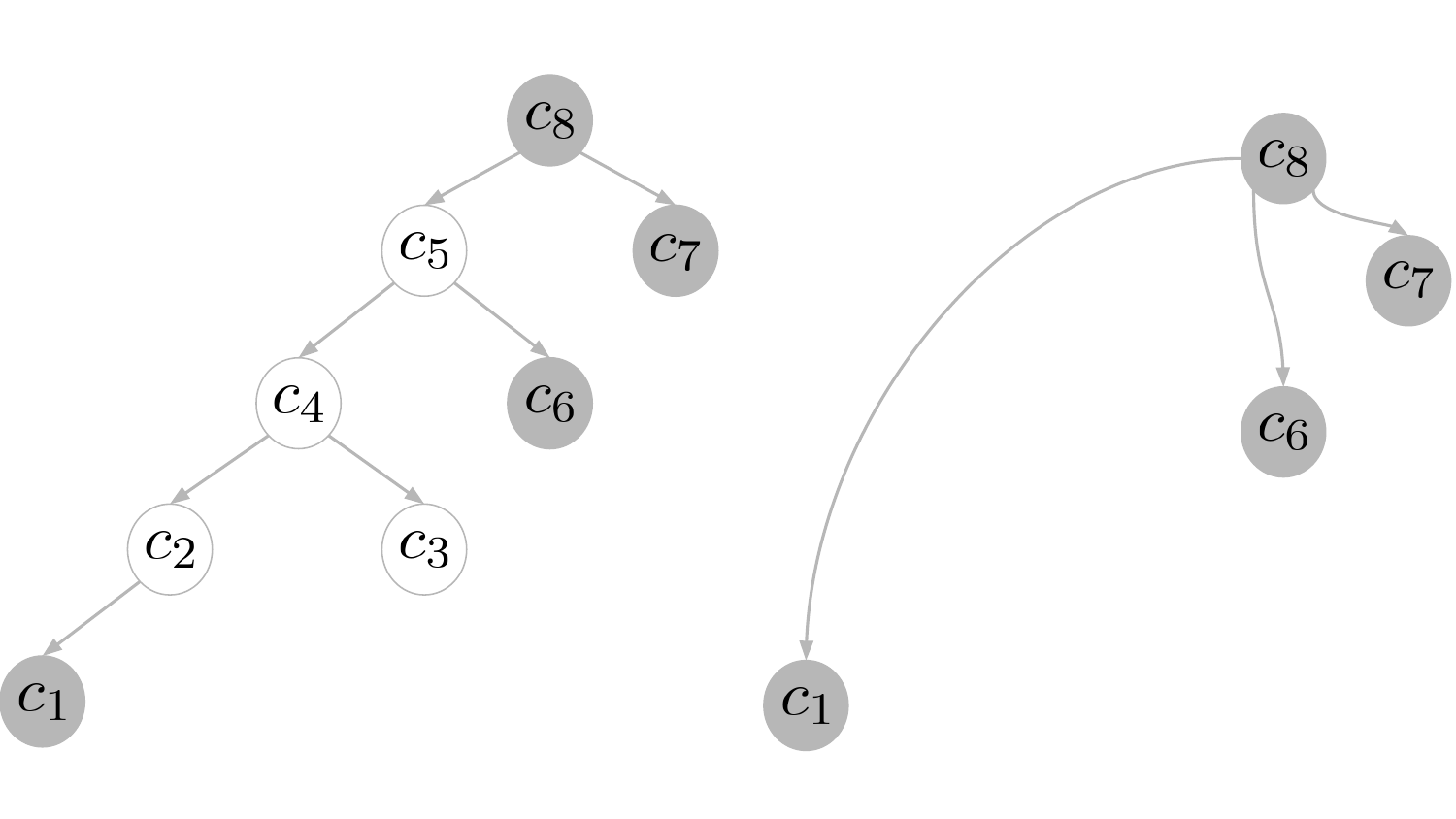}
  \caption{To keep only the selected candidates from the expression forest $\mathcal{F}$ and still preserve the topological order between them, we can represent them as a compressed forest $\mathcal{F}'$. Here, selected candidates are represented by gray nodes, and unselected ones by white nodes. The operation nodes are omitted for simplicity.}
  \label{fig:compression}
\end{figure}

\medskip

As a result, we achieve a linear time complexity instead of exponential, which allows us to perform \textit{more iterations within the same time} and improve the quality of the solution. This result highlights the importance of understanding the interaction of benefits, the selected set of candidates, and the imposed constraints, which we discussed in this paper.

\noindent \textbf{Note on Dimensionality Reduction.} 
Note that \cite{jindal2018selecting} also proposed an additional procedure to reduce the dimensionality of the ILP problem by considering only the variables associated with the currently selected candidates. To show the advantage of our optimization over this technique, we describe a special expression forest \textit{compression} procedure that takes a quadratic time with respect to the number of selected candidates. 

\begin{figure}
  \centering
  \includegraphics[width=0.6\linewidth,keepaspectratio]{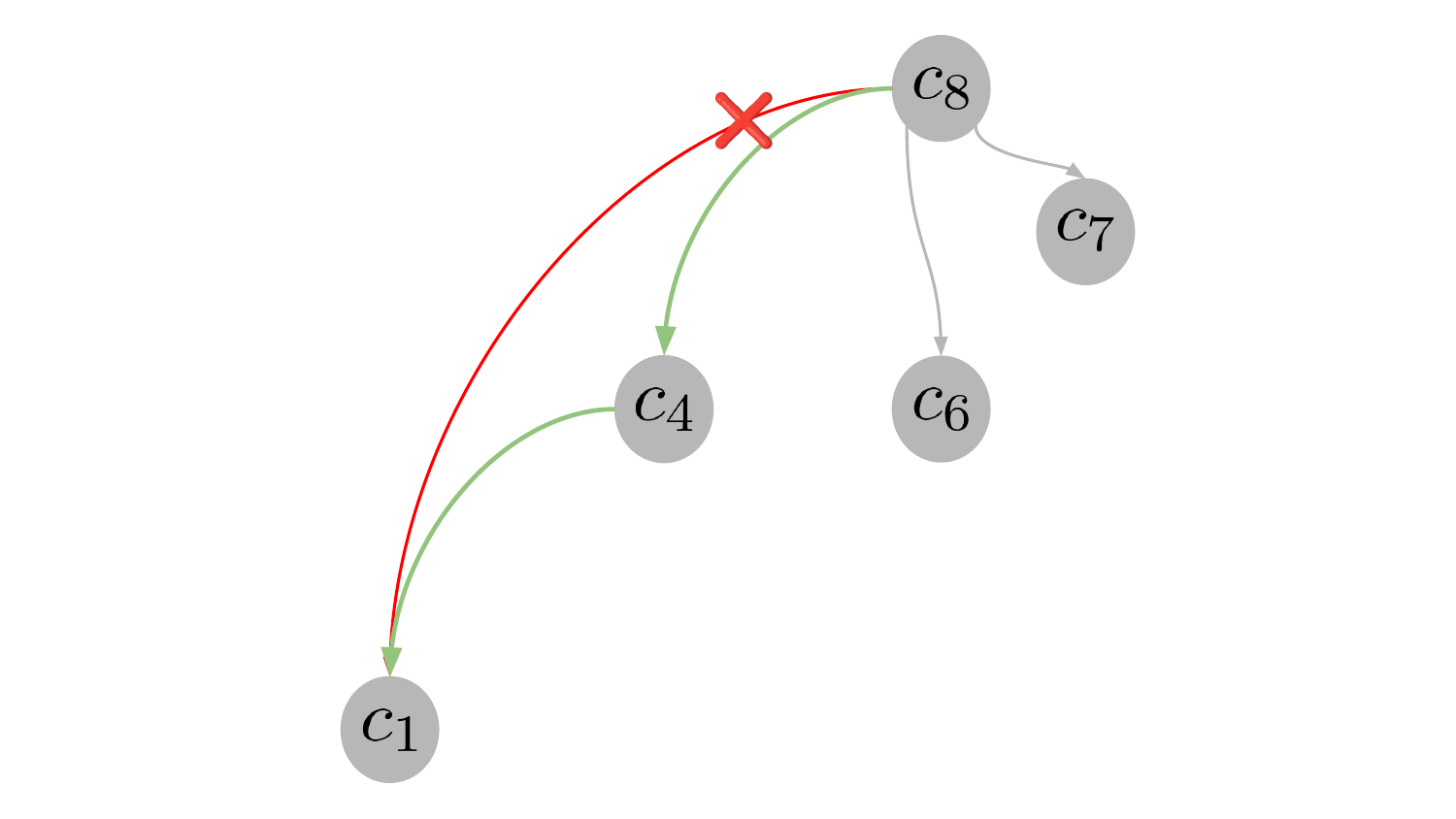}
  \caption{In order to add a new candidate \(c_4\) to the set of selected candidates \(\{c_1, c_6, c_7, c_8\}\), the compressed forest needs to be rebuilt. To do this, we first find the parents of $c_4$ which involves searching for the lowest vertices \(x\) such that \(is\_parent(x,c_4)=True\) (node \(c_8\)). After that, we need to adjust the edges of all their children \(y\) that satisfy the equation \(is\_parent(c_4, y)=True\) (node \(c_1\)).}
  \label{fig:rebuilding}
\end{figure}

The key observation is that the only requirement to the query representation used in our optimization is that the topological order of the candidates must be \textit{maintained}. We show that we can compress the expression forest to include only the current set of selected candidates while preserving their topological order. An example of compression is depicted in Figure \ref{fig:compression}. The only difficulty is that the set of selected candidates \textit{changes} at each iteration, so we also need to \textit{update} the compressed forest. The case of removing a candidate is straightforward: we simply move the edges from the removed vertex to its parent. Adding a new candidate, however, requires to consider the topology induced by the complete expression forest, which is not preserved by compression. A pivotal observation for addressing this issue is that an edge \(parent \rightarrow child\) in the compressed forest exists only if \(parent\) is the lowest ancestor of \(child\) among all selected candidates. Moreover, we need to add the edge \(c_{new\_child} \rightarrow c_{old\_child}\) only to those vertices that either lack a parent or whose lowest ancestor is also the lowest ancestor to \(c_{new\_child}\). The relation \(is\_parent(\cdot, \cdot)\) can be precomputed in quadratic time (in the size of the \textit{original} expression forest), so that a check $is\_parent(c, c')$ for any pair $c, c'$ can be made in \(O(1)\) time. Then by traversing the vertices from top to bottom and by checking that \(is\_parent(x, c_{new\_child}) = True\), we can identify all the lowest ancestors \(c_{parent}\) and subsequently all \(c_{old\_child}\) nodes. This procedure is illustrated in Figure \ref{fig:rebuilding}. As a result, the total running time of our proposed solution to add or remove \textit{one} candidate is linear in the number of selected candidates, i.e., the time for rebuilding the tree is quadratic in the number of candidates.

\begin{mytakeaway}
    Employing the tree structure of query representation for benefit computation allows for exponentially accelerating SotA algorithms, such as RLView and BigSubs, and improving the quality of solution by omitting the constraints on the joint use of nested candidates. 
\end{mytakeaway}

%% file: chapters_challenges.tex
\section{Challenges and Open Problems}
\label{chapter:challenges}

\subsection{Design of Candidates}
\textbf{Candidate Space.} 
No matter what the selection algorithm is, the design of a candidate space must guarantee the existence of good solutions. The question of choosing an appropriate candidate space is highly non-trivial. It has been shown in \cite{chirkova2002formal} that an optimal solution may not be found if one builds a candidate space upon query subexpressions. On the other hand, it is impractical to work with large subexpression spaces. Hence, it is necessary to study ways how to \textit{constrain the candidate space} while providing guarantees on the quality of contained solutions. Heuristic approaches have been already attempted for this purpose (Section \ref{subsubchapter:heuristics_space_reduction}), but we believe that there is much room for new results and advanced techniques in this area.

\noindent \textbf{Computational Complexity.} The formal complexity analysis has made a significant contribution to the development of selection algorithms (Section \ref{subchapter:constraints}). Based on the state-of-the-art results we believe it is worth studying the approximability of the Selection Problem for AND-OR-DAGs. Understanding this issue will facilitate building efficient selection algorithms for this expressive representation framework. The computational complexity is also worth investigating for AND-DAG and Data Cube frameworks, since for these representations the question of \mbox{existence} of an exact polynomial solution \cite{gupta2005selection,karloff1999complexity} is still open. We note that there are also hopes for positive computational results on the little explored classes of binary AND-OR-DAGs.

\subsection{Benefit estimation} 
\textbf{Model architecture.} 
Correct estimation of the benefit of candidates is an important ingredient in solving the Selection Problem (Section \ref{subchapter:hybrid}). One way to obtain accurate estimates is to build a special model for benefit prediction. We noted that using the tree structure of queries allows for designing more optimal selection algorithms (Section \ref{subchapter:heuristics}), and we believe this knowledge should be employed in prediction models as well. While searching for new architectures, it is also worth adopting similar solutions from other fields, among which we highlight Tree Convolution networks \cite{tai2015improved} and TreeLSTM modification of recurrent neural networks \cite{tai2015improved}.

\noindent \textbf{Encoding.} 
To built a vectorized representation of a query, it is important to correctly \textit{encode} its keywords and subexpressions. In  \cite{marcus2019neo}, a special encoding technique similar to word2vec \cite{mikolov2013efficient} was proposed which allows for encoding more information and significantly improves prediction accuracy. State-of-the-Art techniques for encoding plans in prediction models typically use one-hot encoding and linear transformations, which leaves space for advanced encoding techniques.

\subsection{Diverse Scenarios}

\textbf{Dynamic.} 
The algorithms considered in this paper use an implicit assumption that candidates persist until the entire given workload is executed. But if a candidate is found to be useful for executing only a part of the workload, it can be replaced at some point \cite{gupta2001query, zhang2003dynamic}. Controlling the order of query execution and dynamically changing the set of selected candidates can significantly increase the overall performance. We think that this feature should become a part of modern selection algorithms for multi-query optimization. First steps in this direction have already been made, for example, in \cite{liang2019opportunistic}. However, the authors considered only a basic scenario, with subexpression subspace residing in memory, and completely omitted buffer pool state and caching effects.

\noindent \textbf{Distributed.} 
Due to the trend towards scalable computing systems, in order to develop efficient solutions to the Selection Problem, it is necessary to take into account data location in benefit modeling, similar to the ideas from  \cite{chaves2009towards}. 

\noindent \textbf{Partial Selection.} 
In some situations one can partially save a candidate, for example, to keep only its most frequently requested part. Then, in processing  a query, most of the data can be retrieved by reusing the candidate, while the rest of the data can be computed from base relations \cite{luo2006partial}. A similar approach was implemented by using \textit{hotspot} tables in paper \cite{zhou2006dynamic}. We think that these techniques should be further developed, since they enable storing candidates more efficiently and avoiding expensive disk reads.

\subsection{Unified Selection Framework}
\textbf{Interconnections of Problems.} 
The common nature of selection tasks in different domains suggests reuse of ideas. We see that the idea of storing frequent objects from classical caching is adopted in View Selection \cite{dar1996semantic} and Index Selection \cite{seshadri1995generalized}. The reuse of techniques can also be observed in the combined index and view selection algorithms (Sections \ref{subsubchapter:hybrid_early_approaches}, \ref{subsubchapter:heuristics_selection}). We believe there is a strong potential in such approaches. For example, we noticed the non-linear cost of saving in plan caching (see Example \ref{ex:plan_cache}). This leads to the idea to more efficiently store plans by reusing techniques for the maintenance constraint from the field of View Selection. 

\noindent \textbf{Evaluation Platform.} 
With the potential of reuse of techniques between different approaches, we see a great benefit in creating a universal platform for evaluating pros and cons of different selection algorithms. In this paper, we are taking the first step in this direction by proposing a general Candidate Selection framework and by highlighting the common structural properties of candidates from different problems. We are facing however a number of technical challenges along this way, including platform-dependence and closed code of some implementations, as well as the lack of standard benchmarks.

%% file: chapters_conclusion.tex
\section{Conclusion}
\label{chapter:conclusion}
In this paper, we highlighted the cross-domain nature of  selection algorithms and proposed a framework of Candidate Selection which enables the reuse of techniques from these algorithms. The presented framework covers many aspects of Multi-Query Optimization, ranging from View Selection and Index Selection to Query / Plan Caching. We provided a deep analysis on different instances of selection problems and techniques to solve them. As an example application of our study, we showed how to exponentially accelerate some of the SotA View Selection algorithms. We provided a modern classification of selection algorithms, we formulated open issues and challenges, and summarized promising directions for future research. We believe that this survey will aid researchers and practitioners to gain insights on the advantages and shortcomings of existing approaches and to develop novel techniques for Multi-Query Optimization.